\documentclass[sn-mathphys]{sn-jnl}
\usepackage{comment}
\usepackage{dblfloatfix}
\usepackage{algpseudocode} 
\usepackage{graphicx}
\usepackage{tabularx}

\usepackage{amsthm}
\usepackage{url}
\usepackage{bm}

\usepackage{wrapfig}
\usepackage{textcomp}
\usepackage{xcolor}
\usepackage{caption}
\usepackage{subcaption}
\usepackage{comment}
\usepackage{mathtools}
\usepackage{xparse}

\usepackage{arydshln}
\usepackage{amssymb}
\usepackage{pifont}
\newcommand{\cmark}{\ding{51}}%
\usepackage{xcolor}
\usepackage{tablefootnote}
\usepackage{bm, xparse, xifthen}

\usepackage{color,soul}
\usepackage{lipsum}

\DeclareMathOperator*{\argmin}{arg\,min}
\DeclarePairedDelimiter{\norm}{\lVert}{\rVert}

\NewDocumentCommand{\vect}{ O{} O{} m }{\bm{#3}\ifthenelse{\isempty{#1}}{}{^{(#1)}}\ifthenelse{\isempty{#2}}{}{_{#2}}}
\NewDocumentCommand{\mat}{ O{} O{} m }{\bm{#3}\ifthenelse{\isempty{#1}}{}{^{(#1)}}\ifthenelse{\isempty{#2}}{}{_{#2}}}
\NewDocumentCommand{\ten}{ O{} O{} m }{\bm{\mathcal{#3}}\ifthenelse{\isempty{#1}}{}{^{(#1)}}\ifthenelse{\isempty{#2}}{}{_{#2}}}

\jyear{2021}%

\theoremstyle{thmstyleone}%
\theoremstyle{thmstyletwo}%

\theoremstyle{thmstylethree}%

\begin{document}

\title[Distributed Out-of-Memory NMF on CPU/GPU Architectures]{Distributed Out-of-Memory NMF on CPU/GPU Architectures}

\author*[1]{\fnm{Ismael} \sur{Boureima}}\email{iboureima@lanl.gov}

\author[1]{\fnm{Manish} \sur{Bhattarai}}\email{ceodspspectrum@lanl.gov}

\author[1]{\fnm{Maksim} \sur{Eren}}\email{maksim@lanl.gov}
\author[2]{\fnm{Erik} \sur{Skau}}\email{ewskau@lanl.gov}
\author[3]{\fnm{Philip} \sur{Romero}}\email{prr@lanl.gov}
\author[2]{\fnm{Stephan} \sur{Eidenbenz}}\email{eidenben@lanl.gov}
\author[1]{\fnm{Boian} \sur{Alexandrov}}\email{boian@lanl.gov}

\affil[1]{\orgdiv{Theoritical Divison}, \orgname{Los Alamos National Laboratory}, \orgaddress{\city{Los Alamos}, \postcode{87545}, \state{NM}, \country{USA}}}
\affil[2]{\orgdiv{Computer, Computational, and Statistical Science Division}, \orgname{Los Alamos National Laboratory}, \orgaddress{\city{Los Alamos}, \postcode{87545}, \state{NM}, \country{USA}}}
\affil[3]{\orgdiv{HPC Divison}, \orgname{Los Alamos National Laboratory}, \orgaddress{\city{Los Alamos}, \postcode{87545}, \state{NM}, \country{USA}}}


\abstract{We propose an efficient distributed out-of-memory implementation of the Non-negative Matrix Factorization (NMF) algorithm for heterogeneous high-performance-computing (HPC) systems. The proposed implementation is based on prior work on NMFk, which can perform automatic model selection and extract latent variables and patterns from data. In this work, we extend NMFk by adding support for dense and sparse matrix operation on multi-node, multi-GPU systems. The resulting algorithm is optimized for out-of-memory (OOM) problems where the memory required to factorize a given matrix is greater than the available GPU memory. Memory complexity is reduced by batching/tiling strategies, and sparse and dense matrix operations are significantly accelerated with GPU cores (or tensor cores when available). Input/Output (I/O) latency associated with batch copies between host and device is hidden using CUDA streams to overlap data transfers and compute asynchronously, and latency associated with collective communications (both intra-node and inter-node) is reduced using optimized NVIDIA Collective Communication Library (\textit{NCCL}) based communicators. Benchmark results show significant improvement, from 32X to 76x speedup, with the new implementation using GPUs over the CPU-based NMFk. Good weak scaling was demonstrated on up to 4096 multi-GPU cluster nodes with approximately 25,000 GPUs when decomposing a dense 340 Terabyte-size matrix and an 11 Exabyte-size sparse matrix of density $10^{-6}$. 
}

\keywords{NMF, out-of-memory,  latent features , model selection , distributed processing , parallel programming , big data , heterogeneous computing , GPU , CUDA , NCCL , cupy}



\maketitle

\section{Introduction}
NMF is a popular unsupervised learning method that extracts sparse and explainable latent features \cite{lee1999learning}, which are often used to reveal explainable low-dimensional hidden structures that represent and classify the elements of the whole dataset \cite{cichocki2009nonnegative}. NMF is used in big data analysis, which plays a crucial role in many problems, including human health, cyber security, economic stability, emergency response, and scientific discovery. With the increased accessibility to data and technology,  datasets continue to grow in size and complexity. At the same time, the operational value of the information hidden in patterns in such datasets continues to grow in significance. Extracting explainable hidden features from large datasets, collected experimentally or computer-generated, is vital because the data presumably carries essential (but often previously unknown) information about the investigated phenomenon's causality, relationships, and mechanisms. Discovering meaningful hidden patterns from data is not a trivial task because the datasets are formed only by directly observable quantities while the underlying processes or features, in general, remain unobserved, latent, or hidden \cite{everett2013introduction}. 

Analysis of vast amounts of (usually sparse) data via NMF requires novel distributed approaches for reducing computational complexity, speeding up the computation, and dealing with data storage and data movement challenges. Most NMF computations are matrix-matrix multiplications, which GPU accelerators can speed up. The primary performance and scaling limiting factors in NMF implementations on modern heterogeneous HPC systems are high communication costs due to data movement across different system parts (inter-node and intra-node communications). In various cases, these communication delays exceed the time the actual computations take, resulting in poor performance and poor scalability on large distributed systems. 

The growth in data volumes outpacing the improvement in hardware specifications is causing significant challenges in extracting useful information from large-scale datasets using algorithms like NMF. This motivates the need for out-of-memory implementations of NMF for distributed HPC systems, which will allow the decomposition of large datasets that does not fit in memory at once. Enabling out-of-memory factorization is very important because it removes the matrix size constraint imposed by the GPU memory, thus enabling the analysis of datasets up to the cumulative size of all RAM on the cluster. This is mainly required to address the challenges presented by the need to factorize the ever-growing datasets. We utilize this unique ability of pyDNMF-GPU to demonstrate the decomposition of record large dense, and sparse datasets.

To illustrate how \textit{pyDNMF-GPU} can be used as a building block for more comprehensive workflows, we integrate \textit{pyDNMF-GPU} with our existing model selection algorithm 
  \textit{pyDNMFk}\footnote[1]{\textit{pyDNMFk: }\url{https://github.com/lanl/pyDNMFk}} that enables automatic determination of the (usually unknown) number of latent features on a large scale datasets~\cite{alexandrov2013deciphering,alexandrov2020source,chennupati2020distributed,bhattarai2021pydnmfk,vangara2021finding}. We utilized the integrated model selection algorithm  previously to decompose the worlds' largest collection of human cancer genomes ~\cite{alexandrov2013signatures}, defining cancer mutational signatures ~\cite{alexandrov2020repertoire}, as well as successfully applied to solve real-world problems in various fields~\cite{vangara2020semantic,bhattarai2020distributed,alexandrov2019nonnegative,s.20211055,bhattarai2022distributed,pyDRESCALk,vangara2021finding,eren2022general,eren2022fedsplit,eren2022senmfk}.

This integration results in our out-of-memory scalable tool, \textit{pyDNMFk-GPU}, to be capable of estimating the number of latent features in extra-large \emph{sparse} (tens of EBs) and dense (hundreds of TBs) datasets while operating across CPU-GPU hardware. To the best of our knowledge, our framework is the first to identify hidden features in large-scale dense and sparse datasets.

In experiments on large HPC clusters, we show \textit{pyDNMF-GPU}'s potential: we measure up to 76x improvement on a single GPU over running on a single 18-core CPU. We also demonstrate weak scaling on up to 4096 multi-GPU cluster nodes with approximately 25,000 GPUs when decomposing a dense 340 Terabyte-size matrix and an 11 Exabyte-size sparse matrix of density $10^{-6}$.

Our main contribution is a novel NMF parallel framework, called \textit{pyDNMF-GPU}, that minimizes the data movement on GPUs, improving overall running times. 
Our work's main contribution and novelty is the proposal of a new distributed implementation of NMF with low memory complexity that enables the out-of-memory factorization of very large datasets. Our proposed implementation,  \textit{pyDNMF-GPU}, takes advantage of the following three modern design choices: 
\begin{itemize}
\item \textit{pyDNMF-GPU} reduces the latency associated with local data transfer between the GPU and host (and vice-versa) by using \textit{CUDA streams}.
\item Latency associated with collective communications (intra-node and inter-node) is reduced by using \textit{NCCL primitives}.\footnote[2]{\textit{NCCL: }\url{https://developer.nvidia.com/nccl}}.
\item We incorporate a batching approach for inter-node communication, which provides a unique ability to perform out-of-memory NMF while using multiple GPUs for the bulk of computations. 
\end{itemize} 

The main contributions of the paper include:

\begin{itemize}
    \item Introducing a novel distributed algorithm with out-of-memory support for NMF for sparse and dense matrices operating across CPU-GPU hardware.

    \item Report, the first NCCL communicator accelerated NMF decomposition tool in distributed  GPUs. 
  
    \item Demonstrate the framework's scalability over a record-breaking 340 Terabytes (TB) dense and 11 Exabytes (EB) sparse synthetic datasets.

\end{itemize}

The remainder of the paper is organized as follows: Section~\ref{sec:Background} gives a summary of NMF and the existing parallel NMF implementations. In Section~\ref{sec:Implementation}, we detail the design considerations and choices for a scalable, parallel, and efficient algorithm in different configurations of the data size and available GPU VRAM, as well as the complexity of the new implementation. The efficacy of pyDNMF-GPU with different benchmark results and the validation of benchmark results on a synthetic dataset with a predetermined number of latent features is shown in Section~\ref{sec:results}. We finally conclude with summaries and suggestions of possible future work directions in Section~\ref{sec:Conclusion}. 

\section{Background and related work}
\label{sec:Background}
\subsection{Non-negative matrix factorization algorithms}
NMF \cite{lee1999learning} approximates the non-negative observational matrix $\mat{A}\in \mathbb{R}_{+}^{m \times n}$ with a product of two non-negative factor matrices  $\mat{W}\in \mathbb{R}_{+}^{m \times k}$ and $\mat{H}\in \mathbb{R}_{+}^{k \times n}$ where the columns of $\mat{W}$ represent the latent features, while the columns of $\mat{H}$ are the coordinates/weights of the analyzed samples (the columns of $\mat{A}$) in the reduced latent space, and $k$ is the latent dimension of the data. The NMF minimization is based on alternating update of each one of these two factor matrices until convergence indicated by the condition $\Arrowvert \mat{A} - \mat{W} \mat{H} \Arrowvert_F \leq \eta$ is reached. Here $\Arrowvert . \Arrowvert_F$ is the Frobenius norm, $\Arrowvert \mat{A} \Arrowvert_F = \sqrt{\sum_i \sum_j a_{ij}^2}$, where $a_{ij}$ is the element on row $i$ and column $j$, and $\eta$ is the desired tolerance. Each iteration consists of a $\mat{W}$-update sub-step followed by a $\mat{H}$-update sub-step, given by

\begin{equation}
  \begin{aligned}
    \mat{W} &\leftarrow \argmin_{\mat{W} \geqslant 0} \norm{\mat{A} - \mat{W}\mat{H}}_{F}^{2} \\
    \mat{H} &\leftarrow \argmin_{\mat{H} \geqslant 0} \norm{\mat{A} - \mat{W}\mat{H}}_{F}^{2},
  \end{aligned}
  \label{eq:au_update}
\end{equation}

\begin{algorithm}
\scriptsize
    \caption{$\mat{W},\mat{H}= \operatorname{NMF}(\mat{X},k)$ -- Generic {\em  NMF}}
    \label{alg:nmf}
   
    \begin{algorithmic}[1]
     \Require  $\mat{X} \in \mathbb{R}_{+}^{m \times n}$, $k$ is the rank of approximation and $max\_iters$ is the number of iterations.

    \State Initialize $\mat{W}$,$\mat{H}$ = $\operatorname{rand}(m,k)$,$\operatorname{rand}(k,n)$ 
    \State $i = 0$
    \State $\eta_i = \eta$ + 1  \ \ \ \ \ \ \ \ \ \ \ \ \ \ \ \ \Comment{ Ensure $\eta_i > \eta$ to enter loop}         \label{line:init_epsilone}
    \While{ ($\eta_i \geq \eta$ or $i \leq i_{max}$)  }
    \Statex $@$ stands for matrix multiplication operation
    \State    $\mat{W} \leftarrow \mat{W} * \frac{(\mat{A} @ \mat{H}^T)}{\mat{W} @\mat{H} @ \mat{H}+ \epsilon}$  \ \ \ \ \ \ \ \ \ \ \ \ \ \ \ \ \ \ \ \ \ \ \ \ \ \ \  \Comment{$\mat{W}$ update
    } \label{line:W_up} 
    \State    $\mat{H} \leftarrow \mat{H} * \frac{(\mat{W}^T @ \mat{A})}{\mat{W}^T @\mat{W} @ \mat{H} + \epsilon}$  \ \ \ \ \ \ \ \ \ \ \ \ \ \ \ \ \ \ \ \ \ \ \ \ \ \ \Comment{$\mat{H}$ update} \label{line:H_up} \; \\
    \State    $\mat{X} \leftarrow  \mat{W} \mat{H}$ \label{X_reconn}
    \State    $\eta_i = \Arrowvert \mat{A} - \mat{X}\Arrowvert_F^2$ \label{NMF_norm} 
    \State    $i = i+1$ ;
    \EndWhile
    
    \end{algorithmic}
    \end{algorithm}

The Frobenius norm (FRO) based multiplicative update (MU) algorithm is presented in Algorithm~\ref{eq:au_update}. 
In addition to the presented Frobenius norm-based MU algorithm (which leads to a Gaussian model of the noise ~\cite{fevotte2009nonnegative}) other similarities (e.g., KL-divergence that corresponds to a Poisson model) can also be used in the NMF minimization. Also, based on the update rules, several variants of NMF algorithms exist such as Hierarchical Alternating Least Squares (HALS)~\cite{phan2008multi}, Alternating Non-negative Least Squares with Block Principle Pivoting (ANLS-BPP)~\cite{kim2012fast}, and Block coordinate descent algorithm (BCD)~\cite{kim2014algorithms}. These algorithms have different advantages in the context of convergence rate, computational, and memory requirements. MU-based updates are computationally and memory-wise cheap at the cost of slower convergence. Whereas HALS, BCD, and ANLS-BPP have faster convergence rates at the cost of higher computational and memory requirements and high communication costs for parallel implementations. 
In our experiments, we use the FRO-based MU algorithm to demonstrate record scalability on large datasets due to its lower computation and communication cost, which can easily be modified with another update algorithm or similarity metric.

\subsection{Related work on distributed NMF}

Several parallel implementations have been proposed to address the computational need of NMF for large datasets involving multiple and repeated matrix-matrix multiplications of several orders in magnitude. The existing parallel implementations can be grouped under two categories (i) with shared memory and (ii) with distributed memory. Majority of existing parallel works utilize shared-memory multiprocessor ~\cite{battenberg2009accelerating,fairbanks2015behavioral,moon2020alo,phipps2019software} and shared memory GPUs~\cite{mejia2015nmf,moon2020alo,lopes2010non,phipps2019software} via OpenMP and CUDA libraries respectively. A majority of distributed memory implementations rely on \emph{MPI primitives} for distributed CPU~\cite{bhattarai2020distributed,kannan2016high} and CUDA-aware \emph{MPI primitives} for distributed GPU ~\cite{kannan2016high,mejia2015nmf} parallelization. Although shared-memory implementations drastically minimize the communication costs incurred for distributed memory implementation~\cite{moon2020alo}, there is a constraint on how much data such frameworks can decompose. Due to this constraint, shared-memory implementation often cannot provide the computational/memory requirements needed for the current large-scale datasets. 

\begin{table}[t!]
\caption{\label{tab:gpu-nmfs} A comparison chart for different GPU based NMF implementations.}
\resizebox{\columnwidth}{!}{
\centering
\begin{tabular}{|c|c|c|c|c|c|}
\hline
     \textbf{Framework} & \textbf{GPU} & \textbf{Multi-GPU} & \textbf{Sparse} & \textbf{Out-of-Memory} & \textbf{Remarks} \\
      \hline
    
     \textit{nmf-cuda}\cite{battenberg2009accelerating} & \cmark & -- & -- & -- & Multithreading support only\\
\hdashline
      &&&&& Communication inefficient \\
     \textit{NMF-mGPU}\cite{mejia2015nmf}&\cmark & \cmark & -- & -- &design,entire factor need \\
     &&&&& to be stored for each GPUs. \\
\hdashline
 \textit{nmfgpu4R}\cite{koitka2016nmfgpu4r} & \cmark & -- & -- & -- & No distributed support. \\
\hdashline
 \textit{pytorch-NMF} & \cmark & -- &\cmark & -- & No distributed support. \\
\hdashline
\textit{NMF-spark}\cite{tang2021collaborative} & \cmark & \cmark & -- & -- & Inefficient scaling results.\\
\hdashline
  &&&&& Shared memory implementation\\
   &&&&&of optimized HALS algorithm \\
 \textit{ALO-NMF}\cite{moon2020alo} &\cmark & -- & \cmark & -- & \&lack of distributed support \\

\hdashline
  &&&&& Not a  significant gain of GPUs\\
   &&&&&over CPUs and lack of \\
 \textit{PLANC}\cite{kannan2016high} &\cmark & \cmark & \cmark & -- & ability to handle large \\
 &&&&& sparse and dense data on GPUs.\\
\hdashline
 
      &&&&& NCCL based   efficient implementation\\
     &&&&& with significant speedup of GPUs  \\
     \textit{pyDNMF-GPU} \textbf{(ours)} & \cmark & \cmark & \cmark & \cmark &  over CPUs and  demonstrated \\
     &&&&&  scaling performance over 340TB \\
      &&&&&  dense and 11EB sparse data.\\
    
\hline
\end{tabular}
}
\vspace{-1.2em}
\end{table}

Almost all distributed GPU implementations including \textit{NMF-mGPU}~\cite{mejia2015nmf} and \textit{PLANC}~\cite{eswar2021planc} rely on significant data communication for the update of the factors. This involves using CUDA-aware MPI primitives for data communication or MPI distributed memory offload through NVBLAS~\cite{eswar2021planc} without multi-node GPU communicators. Such implementation leads to high data movement costs due to data on-loading/offloading to/from the device, which significantly raises communication costs compared to the computation cost for large data decomposition. This is previously illustrated with distributed BPP in \textit{PLANC}~\cite{kannan2016high} and distributed  MU and BCD~\cite{bhattarai2020distributed} where the communication cost is minimized by communicating only with the two-factor matrices and other partitioned matrices among MPI processes. These works attempt to reduce the bandwidth and data latency using  MPI collective communication operations. For distributed CPU implementations, this approach works well as the communication cost is significantly lower compared to the computation cost. However, for GPU implementation, communication cost is higher due to device/host data transfer; therefore, communication cost is a limiting factor for parallel performance when using many GPUs. 

Table~\ref{tab:gpu-nmfs} illustrates the comparison against the existing parallel NMF implementations. Further, support for factorization of sparse datasets equally adds value for our new \emph{pyDNMF-GPU} framework. Since many of the extra-large datasets, such as the text corpora, knowledge graph embeddings (and, in general, most of the relational datasets), cyber network activity datasets, and many others, are highly sparse, having sparse decomposition support dramatically reduces the memory and computational requirements which otherwise would be a major bottleneck for the dense implementation. Despite the support for a sparse dataset for shared-memory in \textit{ALO-NMF} and \emph{genten} ~\cite{moon2020alo,phipps2019software} and for distributed memory in \textit{PLANC}~\cite{kannan2016high}, there is no specific solution aiming to address the bottlenecks due to extracted dense factors and their communications for large sparse datasets. Even though the largest sparse datasets may be a few MBs in size, due to their extreme sparsity, decomposing such datasets would be challenging for most existing frameworks as the extracted factors are dense and very large. Even for such a small non-zero valued size, the corresponding dense factors could easily explode and require an expensive communication of dense intermediate terms. However, our batching framework provides a solution by accommodating larger intermediate-dense factors,  which have not been addressed previously. 
\begin{figure*}[t!]
\centering
    \begin{minipage}[b]{0.46\linewidth}
        \includegraphics[width=1\linewidth]{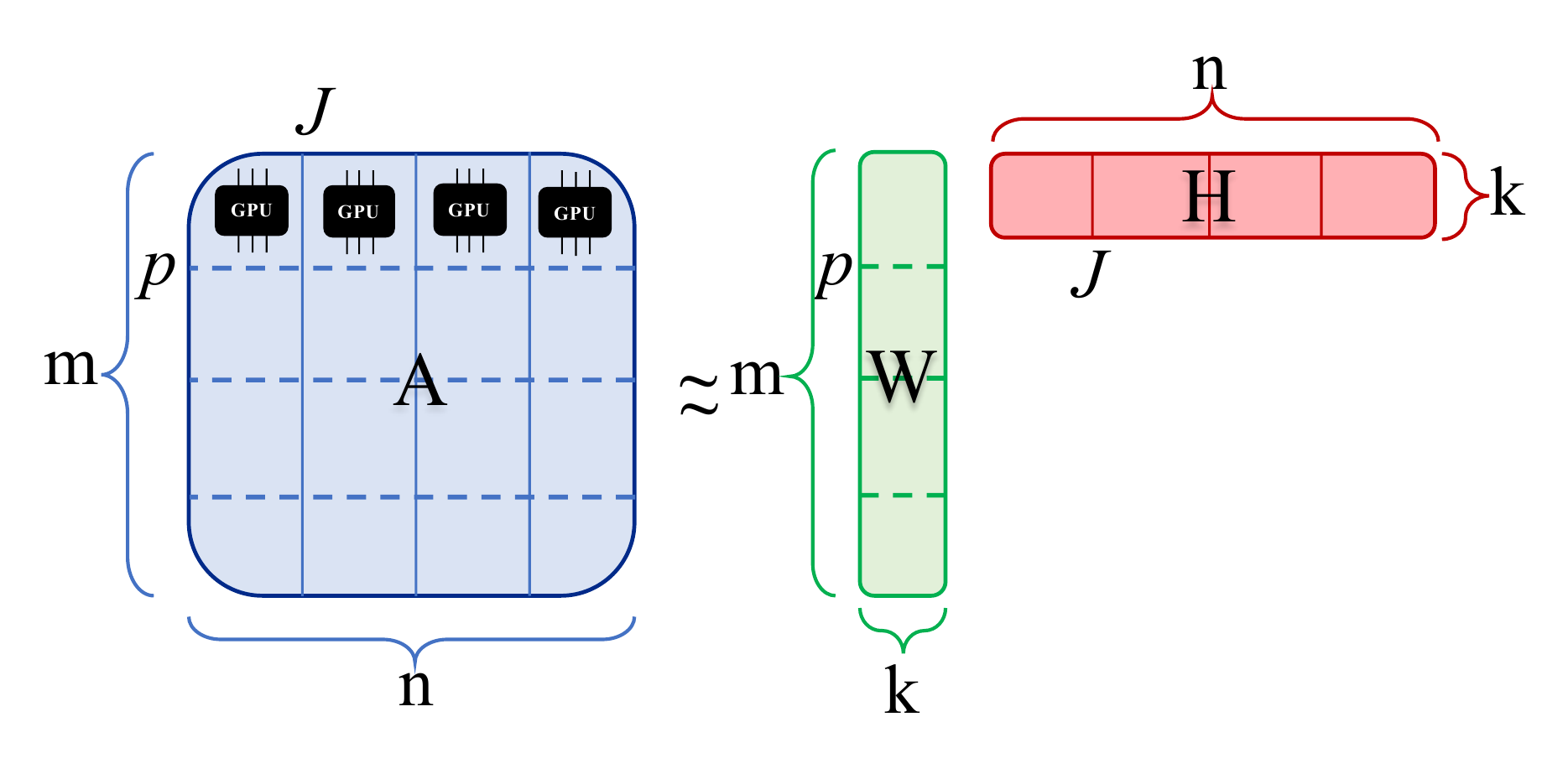}
        \subcaption{Column partition with \emph{orthogonal} batching  \label{fig:nmf_col_part}}
    \end{minipage}
    \quad
    \begin{minipage}[b]{0.46\linewidth}
        \includegraphics[width=1\linewidth]{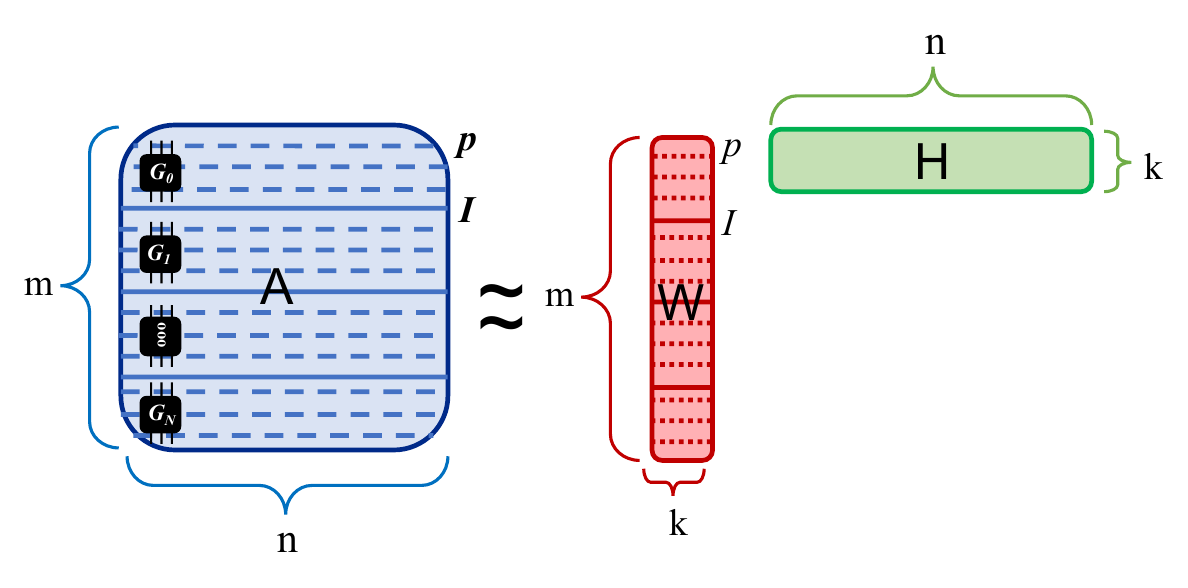}
        \subcaption{Row partition with \emph{co-linear} batching \label{fig:nmf_row_part}}
    \end{minipage}

    \caption{ Illustration of distributed matrix $\mat{A}$ and co-factors $\mat{W}$ and  $\mat{H}$ in \emph{CNMF} and \emph{RNMF} distributed partitions respectively in (a) and (b). Solid lines show distributed partition boundaries, and dashed lines show local partition segmentation in batch for Out-of-Memory decomposition. \label{fig:nmf_distributed_part}} 
    \vspace{-1.3em}
\end{figure*}

\subsection{Rationale for an algorithm for the out-of-memory distributed NMF}

In \emph{pyDNMF-GPU}, we use a distributed implementation of NMF that aims at efficiently factorizing matrices of all sizes, even those too big to fit on available memory, in out-of-memory scenarios. To this end, \emph{pyDNMF-GPU} accelerates matrix operations using GPUs on modern heterogeneous systems, provides support for sparse matrix operations to deal with practical data sets which are often sparse, and can partition large problems into smaller problems solved in a distributed manner. Above all, and to the best of our knowledge, our proposed implementation is the first to provide a solution for practical out-of-memory cases that require the factorization of data too big to be stored on combined available GPU memory.

When performing NMF on GPUs, OOM situations can arise in various scenarios with different degrees of complexity. As discussed in ~\cite{boureima2022distributed}, we distinguish three main types of OOM scenarios. Scenarios of \emph{type 0} (OOM-0) concern practical problems where the input data $\mat{A}$ and its co-factors $\mat{W}$ and $\mat{H}$ can easily be stored on GPU memory. However, an explosion of memory requirement can occur, either due to the unknown rank $k$ becoming significant, causing $\mat{W}$ and $\mat{H}$ to become prohibitively expensive to store on memory, or when computing intermediate results such as $\mat{X}=\mat{W}\mat{H}$ (line~\ref{X_reconn} of Algorithm~\ref{alg:nmf}), when $\mat{A}$ is a large sparse matrix of very low density, where $\mat{X}$ resulting from the operation becomes dense and very likely impossible to store on GPU. For instance, if $\mat{A}\in \mathbb{R}^{10^6 \times 10^6}$ is a sparse matrix, with density of $\delta \approx 10^{-3}$, the size of $\mat{A}$ in dense format, in single precision, is $S_{\mat{A}}\approx 4 TB$, however representing $\mat{A}$ in CSR sparse format can lower the size of $\mat{A}$ down to $S_s \sim 3\times S_A \times \delta \approx 12~GB$ (the factor of 3 accounts for storing the data, indices and index pointers for CSR format), consequently $S_{NMF} \approx 2 \times S_{\mat{A}}\approx 4TB$. Assuming very small k, $\mat{A}$ and all co-factors can be stored on GPU; however, the calculation of the intermediate product $\mat{X}$ from  $\mat{X} = \mat{W}\mat{H}$ would still require a whopping $\sim 8~TB$ of GPU memory (line~\ref{X_reconn} of Algorithm~\ref{alg:nmf}), making this scenario an OOM-0 problem. 
 
 A more complex OOM scenario, \emph{type 1} (OOM-1), arises in cases where matrix $\mat{A}$ and at most one of its co-factors cannot be cached on GPU memory; this is typically the case when dealing with a large $\mat{A}$ that is dense or sparse with high density. Scenarios of \emph{type 2} (OOM-2) are the most complex and consist of practical cases where neither $\mat{A}$, nor its co-factors can be stored on GPU memory. Note that more complexity can arise in cases where data cannot fit on host RAM memory, but that still is of \emph{type 2} as the OOM classification here is based on the GPU RAM memory utilization.
 In other words, in OOM-0 scenarios, all the data can be cached on GPU; in OOM-1 scenarios, the data can partially be cached on GPU, and in OOM-2 scenarios, none of the data can be cached on GPU.
  The treatment of OOM-2 scenarios is out of the scope of this study.

  OOM-0 cases can easily be handled using \emph{tiling} techniques, and OOM-1 cases can be handled with \emph{batching} techniques. In extreme OOM-1 cases, we will complement \emph{batching} by \emph{tiling} to further reduce memory footprint.

 Both \emph{batching} and \emph{tiling} are block-based computational techniques designed to simplify larger, memory-intensive computations into smaller, manageable, and partially solvable tasks. Each technique, however, functions in a distinct setting and serves a different purpose. \emph{Batching} is a process that operates on the host, necessitating consistent data transfer between the host and the device. The efficacy of batching techniques is heavily reliant on the speed of the interconnecting buses between the host and device, such as PCIe or NV-Link. \emph{Batching} techniques become crucial when dealing with OOM-1 problems, as they help in transferring partially computed results. Conversely, \emph{tiling} happens directly within the device memory, resulting in data transfer between global memory and shared or cache memory. The performance of \emph{tiling} techniques is primarily governed by the GPU architecture, including features like memory speed and available shared memory. \emph{Tiling} techniques are especially effective for tackling OOM-0 problems, as they handle computational tasks directly on the device. Notably, \emph{batching} is typically irrelevant for OOM-0 problems as these computations are already based on the device. Similarly, \emph{tiling} techniques alone cannot address OOM-1 issues due to the preliminary need to transfer operands to the device. However, an optimized solution for extreme OOM-1 problems can be achieved by strategically combining both \emph{batching} and \emph{tiling} techniques, thus enhancing the overall performance.

 In the section below, we discuss our implementation and design choices. 

\begin{figure*}[ht!]
\centerline{
\includegraphics[width=.9\linewidth]{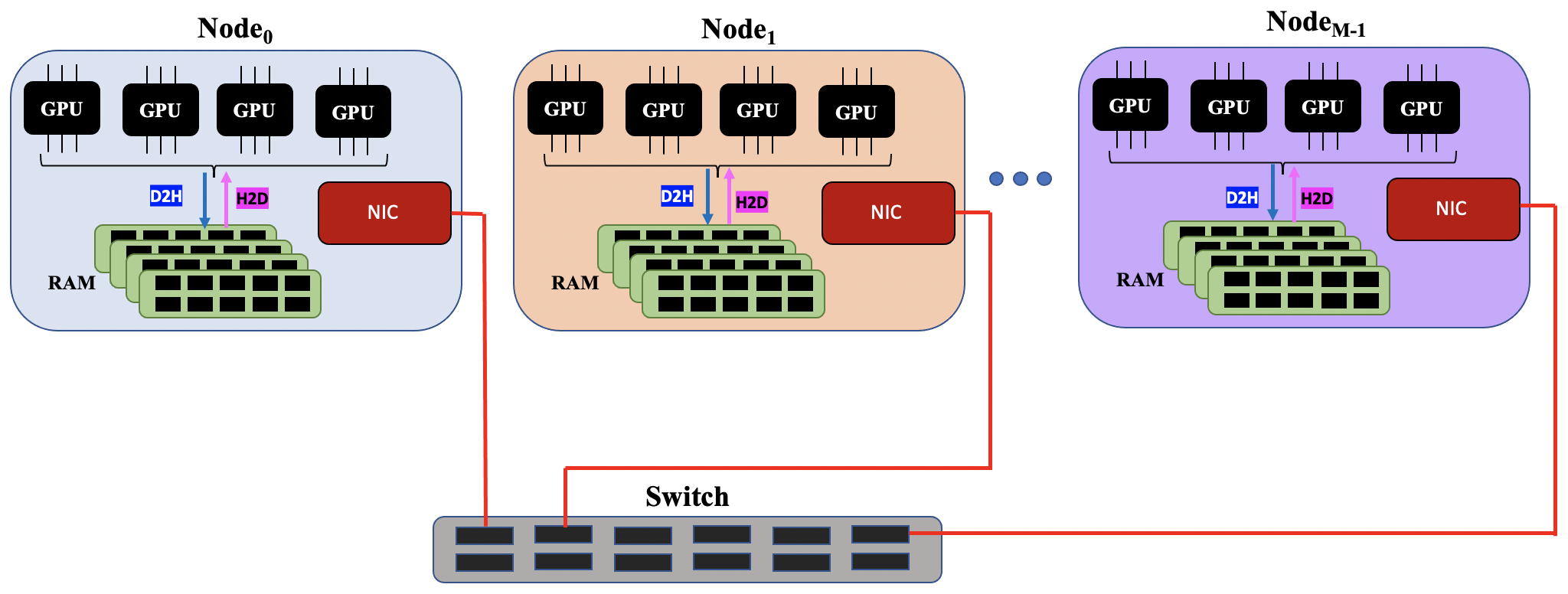}} 
\caption{Illustration of distributed HPC hardware and different communication channels.}
\label{fig:hardware}

\end{figure*}

\section{\emph{pyDNMF-GPU} for heterogeneous systems}
\label{sec:Implementation}

An efficient implementation of NMF for distributed heterogeneous systems should avoid high costs associated with communication (data transfer) resulting from poor consideration for data locality in the distribution of the computational work. Furthermore, cases, where resources such as available combined GPUs memory are limited will require additional considerations and various trade-offs. For instance, it is sometimes better to replicate data over the distributed compute grid to reduce communication. Other times, it is acceptable to use batching techniques that can increase communication costs to lower the memory footprint. Below we first discuss our distributed data partition strategies that partition large problems into smaller problems solvable on cooperative distributed systems in subsection~\ref{subsec:dist_partition}, and then in subsection~\ref{subsec:dist_batching} we discuss our tiling and batching approaches, respectively used to handle practical scenarios of complexities $type 0$ and $type 1$.
\subsection{Distributed implementation}
\label{subsec:dist_partition}

Our implementation considers two one-dimensional data partition strategies based on the shape of matrix $\mat{A}$ ($m\times n$). A column (vertical) partition, CNMF employed when $n > m$, and a row (horizontal) partition, RNMF, is used otherwise.

\begin{algorithm}
\scriptsize
    \caption{$\mat{W},\mat{H}= \operatorname{DCNMF}(\mat{X},k)$--\emph{Distributed CNMF Algorithm}}
    \label{alg:disributed_CNMF}
    \begin{algorithmic}[1]
        \Require  $\mat{X} \in \mathbb{R}_{+}^{m \times n}$, $k$ is the rank of approximation and $max\_iters$ is the number of iterations.
   \Require $\mat{X}$ distributed across $N$ GPUs where  $\mat{X}_{j} \in \mathbb{R}^{m \times n/N}$ where $J=n/N$ if $m<n$. Similarly the co-factor $\mat{W}$ local to each GPU given by $\mat{W}\in \mathbb{R}^{m \times k}$ which is reproduced across different GPUs. $\mat{H}$ is distributed across N GPUS such that $\mat{H}_j \in \mathbb{R}^{k \times J}$. 

    \State Initialize $\mat{W}$,$\mat{H}_j$ = $\operatorname{rand}(m,k)$,$\operatorname{rand}(k,J)$ 

    \Statex $J=n/N, j_0=gID*J, j_1 = (gID+1)*J$
    
  \For {$l=0$ to $l < max\_{iter}$} 
      \State $\mat{WTA} = {\mat[l][]{W}}^T @ \mat{A} $  \label{WTA}   \Comment{$@$ stands for matrix multiplication operation}
        \State $\mat{WTW} =  {\mat[l][]{W}}^T @ \mat{W} $ \label{WTW} 
        \State $\mat[l+1][]{H} = (\mat[l][]{H} * \mat{WTA}) / (\mat{WTW} @ \mat[l]{H} + \epsilon) $ \label{Hupdate} \Comment{$H_{update}$} 

    \State $\mat{HHT} = \mat[l+1][]{H} @\mat{H^{T^{(l+1)}}} $   \label{HHT1}                         
    \State $\mat{HHT} = \operatorname{All\_Reduce}(\mat{HHT})$ \label{HHT}
      \State $\mat{WHHT} = \mat[l]{W} @ \mat{HHT} $ \label{WHHT}
        \State $\mat{AHT} = \mat{A} @ \mat{HT}  $ \label{AHT1} 
        \State $\mat{AHT} = \operatorname{All\_Reduce}(\mat{AHT})$ \label{AHT}
        \State $\mat[l+1][]{W}   = \mat[l][]{W}  * \mat{AHT}/(\mat{WHHT}+\epsilon)$ \Comment{$W_{update}$} \label{Wupdate}
     \EndFor
  
    \end{algorithmic}
\end{algorithm}

\begin{algorithm}
\scriptsize
    \caption{$\mat{W},\mat{H}= \operatorname{DRNMF}(\mat{X},k)$--\emph{Distributed RNMF Algorithm}}
    \label{alg:disributed_RNMF}
    \begin{algorithmic}[1]
        \Require  $\mat{X} \in \mathbb{R}_{+}^{m \times n}$, $k$ is the rank of approximation and $max\_iters$ is the number of iterations.
   \Require $\mat{X}$ distributed across $N$ GPUs where  $\mat{X}_{i} \in \mathbb{R}^{m/N \times n}$ where $I=m/N$ if $n<m$. Similarly the co-factor $\mat{H}$ is reproduced across different GPUs given by $\mat{H}\in \mathbb{R}^{k \times n}$ . $\mat{W}$ is distributed across N GPUS such that $\mat{W}_i \in \mathbb{R}^{I \times k}$.

    \State Initialize $\mat{W}_i$,$\mat{H}$ = $\operatorname{rand}(I,k)$,$\operatorname{rand}(k,n)$ 

    \Statex $I=m/N, i_0=gID*I, i_1 = (gID+1)*I$
    
  \For {$l=0$ to $l < max\_{iter}$} 
      \State $\mat{WTA} = {\mat[l][]{W}}^T @ \mat{A} $  \label{WTA}   \Comment{$@$ stands for matrix multiplication operation}
        \State $\mat{WTA} = \operatorname{All\_Reduce}(\mat{WTA})$ \label{WTA_AllReduce}
        \State $\mat{WTW} =  {\mat[l][]{W}}^T @ \mat{W} $ \label{WTW}
        \State $\mat{WTW} = \operatorname{All\_Reduce}(\mat{WTW})$ \label{WTW_AllReduce}
        \State $\mat[l+1][]{H} = (\mat[l][]{H} * \mat{WTA}) / (\mat{WTW} @ \mat[l]{H} + \epsilon) $ \label{Hupdate} \Comment{$H_{update}$} 

    \State $\mat{HHT} = \mat[l+1][]{H} @\mat{H^{T^{(l+1)}}} $   \label{HHT1}                         
      \State $\mat{WHHT} = \mat[l]{W} @ \mat{HHT} $ \label{WHHT}
        \State $\mat{AHT} = \mat{A} @ \mat{HT}  $ \label{AHT1} 
        \State $\mat[l+1][]{W}   = \mat[l][]{W}  * \mat{AHT}/(\mat{WHHT}+\epsilon)$ \Comment{$W_{update}$} \label{Wupdate}
     \EndFor
  
    \end{algorithmic}
\end{algorithm}

Assuming a distributed system with $N$ GPUs where each GPU is indexed by its global rank $g_{ID}$. In the \emph{CNMF} approach illustrated in Figure~\ref{fig:nmf_col_part}, the $j^{th}$ GPU with $g_{ID}=j$ will work on array partitions $\mat{A}[:, j_0:j_1]$, $\mat{H}[:, j_0:j_1]$ and $\mat{W}$, where $j_0=j \times J$, $j_1=(j+1) \times q$, and $J=n/N$(partition size). Each GPU gets a full copy of $\mat{W}$ ($\mat{W}$ is replicated) and a unique partition of $\mat{A}$ and $\mat{H}$. This translates into a segmentation of arrays $\mat{A}$ and $\mat{H}$ on global memory illustrated with solid lines in Figure~\ref{fig:nmf_col_part}. These solid lines indicate boundaries in global memory and consequently help conceptualize where communication is required whenever information is exchanged from one bounded region to another. The $H$-update is embarrassingly parallel since $\mat{W}^T\mat{W}$, $(\mat{W}^T\mat{W})\mat{H}$, and $\mat{W}^T\mat{A}$ can all be computed locally on each GPU; the $\mat{W}$-update on the other hand, will require two separate all-reduce-sum communications to compute $\mat{A}\mat{H}^T$ and $\mat{H}\mat{H}^T$ as indicated in Algorithm~\ref{alg:disributed_CNMF} lines~\ref{AHT} and\ref{HHT}. 

Following a similar analogy, a \emph{RNMF} approach results with $\mat{H}$ replicated on the different GPUs and $\mat{A}$ and $\mat{H}$ distributed across the compute grid. This time $\mat{W}$-update is embarrassingly parallel since $\mat{H}\mat{H}^T$, $\mat{W}(\mat{H}\mat{H}^T)$, and $\mat{A}\mat{H}^T$ can all be computed locally on each GPU, but the $\mat{H}$-update will require separate all-reduce-sum communication to compute $\mat{W}^T\mat{W}$ and $\mat{W}^T\mat{A}$ as presented in Algorithm~\ref{alg:disributed_RNMF} .

Communication takes place through various channels with different bandwidths and latency. We refer to intra-node communications as any communication on the same node, i.e., yellow, pink, and black lines in Figure~\ref{fig:hardware} and those between different nodes as inter-node communications. i.e red lines in Figure~\ref{fig:hardware}. The latter often have the lowest bandwidth and highest latency and could easily cause bottlenecks for distributed algorithms such as NMF. For these practical reasons, in our implementation, we avoid all-reduce collective calls as much as possible. 
When $n>m$, \emph{CNMF} is more efficient than \emph{RNMF} because it costs less to communicate $\mat{A}\mat{H}^T$ of shape $m \times k$, and  \emph{RNMF} is more efficient when $m>n$ because it cost less to communicate $\mat{W}^T\mat{A}$ of shape $k \times n$.

The FLOP (Floating Point Operations) count for the given Distributed RNMF (Row-wise Nonnegative Matrix Factorization) algorithm can be calculated by going through each of the operations performed in the algorithm. Below is a rough estimation of the FLOP count for each line of interest in the algorithm:

\begin{itemize}

\item  Matrix Multiplication (Line \ref{WTA}): $\mat{WTA} = {\mat[l][]{W}}^T @ \mat{A} $. Here we have a matrix multiplication of size $(k \times I) * (I \times n)$, which will result in $2k*I*n - k*n$ FLOPs.

\item  Matrix Multiplication (Line \ref{WTW}): $\mat{WTW} =  {\mat[l][]{W}}^T @ \mat{W} $. Here we have a matrix multiplication of size $(k \times I) * (I \times k)$, which will result in $2k*I*k - k*k$ FLOPs.

\item  Elementwise Multiplication and Division (Line \ref{Hupdate}): $\mat[l+1][]{H} = (\mat[l][]{H} * \mat{WTA}) / (\mat{WTW} @ \mat[l]{H} + \epsilon) $.This consists of $k*n$ FLOPs for elementwise multiplication and $k*n$ FLOPs for elementwise division, so total $2*k*n$ FLOPs.

\item  Matrix Multiplication (Line \ref{HHT1}): $\mat{HHT} = \mat[l+1][]{H} @\mat{H^{T^{(l+1)}}} $.Here we have a matrix multiplication of size $(k \times n) * (n \times k)$, which will result in $2k*n*k - k*k$ FLOPs.

\item  Matrix Multiplication (Line \ref{WHHT}): $\mat{WHHT} = \mat[l]{W} @ \mat{HHT} $. Here we have a matrix multiplication of size $(I \times k) * (k \times k)$, which will result in $2I*k*k - I*k$ FLOPs.

\item  Matrix Multiplication (Line \ref{AHT1}): $\mat{AHT} = \mat{A} @ \mat{HT}  $. Here we have a matrix multiplication of size $(I \times n) * (n \times k)$, which will result in $2I*n*k - I*k$ FLOPs.

\item  Elementwise Multiplication and Division (Line \ref{Wupdate}): $\mat[l+1][]{W}   = \mat[l][]{W}  * \mat{AHT}/(\mat{WHHT}+\epsilon)$. This consists of $I*k$ FLOPs for elementwise multiplication and $I*k$ FLOPs for elementwise division, so total $2*I*k$ FLOPs.
\end{itemize}

Note: The All\_Reduce operation (Lines \ref{WTA_AllReduce} and \ref{WTW_AllReduce}) are communication operations and are not considered in the FLOP count as they do not involve any computation.

So, total FLOPs for each iteration of the loop = $2k*I*n + 2k*I*k + 2*k*n + 2k*n*k + 2I*k*k + 2I*n*k + 2*I*k - k*n - k*k - k*k - I*k - I*k$.

For $max_{iter}$ iterations, total FLOPs would be $max_{iter}$ times the FLOPs per iteration. Now, to compute GFLOPS, we have GFLOPS = total\_FLOPs/(total\_time$\times 1e9$). Morever, given device peak GFLOPS (peakG), we can compute efficiency as GFLOPS/peakG*100\%.


The total VRAM required to factorize $\mat{A}$ of size $size(\mat{A}) = S_A$ (in Bytes) is typically in the order of $S_{NMF} \sim 4 \times S_A$. One fold of $S_A$ to store $\mat{A}$ in memory, another fold to store perturbed $\mat{A}$\cite{bhattarai2021pydnmfk}, an additional fold to compute intermediate product $\mat{X}=\mat{W}@\mat{H}$ when checking the convergence condition $\Arrowvert \mat{A} - \mat{W} \mat{H} \Arrowvert_F \leq \eta$, and almost one full fold to store the co-factors $\mat{W}$, $\mat{H}$, and heavy intermediate products such as $\mat{W}^{T}\mat{A}$ or $\mat{A}\mat{H}^{T}$. When the total available combined GPU VRAM, $S_{GV}$, is lower than $S_{NMF}$, as in practical big data applications, batching techniques are imperative. The batching, in most cases, increase intra-node and inter-node communication overheads. Although this can significantly affect the algorithm's performance, proper use of asynchronous data copy and CUDA streams can reduce performance loss by overlapping compute and data transfers, as discussed in our out-of-memory implementation below.

\begin{figure*}[t!]
\centerline{
\includegraphics[width=.9\linewidth]{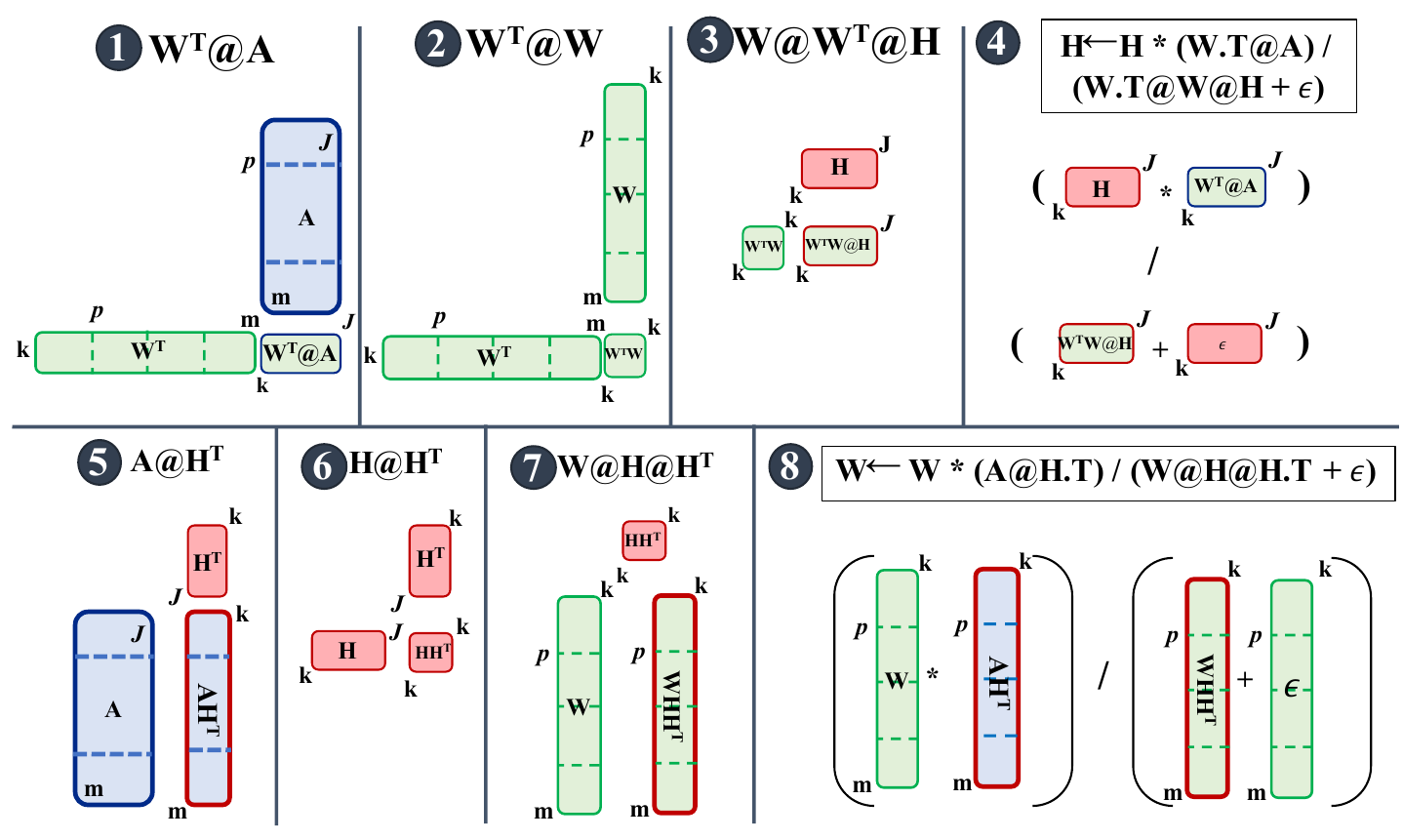}} 
\caption{Illustration of the batched multiplicative update of Algorithm~\ref{alg:batched_OCNMF} for the column partition(CNMF). Green array is duplicated across different MPI ranks. Blue and red arrays are distributed, and only red array is cached on device. For CNMF, p is Out-of-Memory batch width and J is distributed partition width. \label{fig:MU_batch_col_algo}}
\end{figure*}

\begin{figure*}[t!]
\centerline{
\includegraphics[width=.9\linewidth]{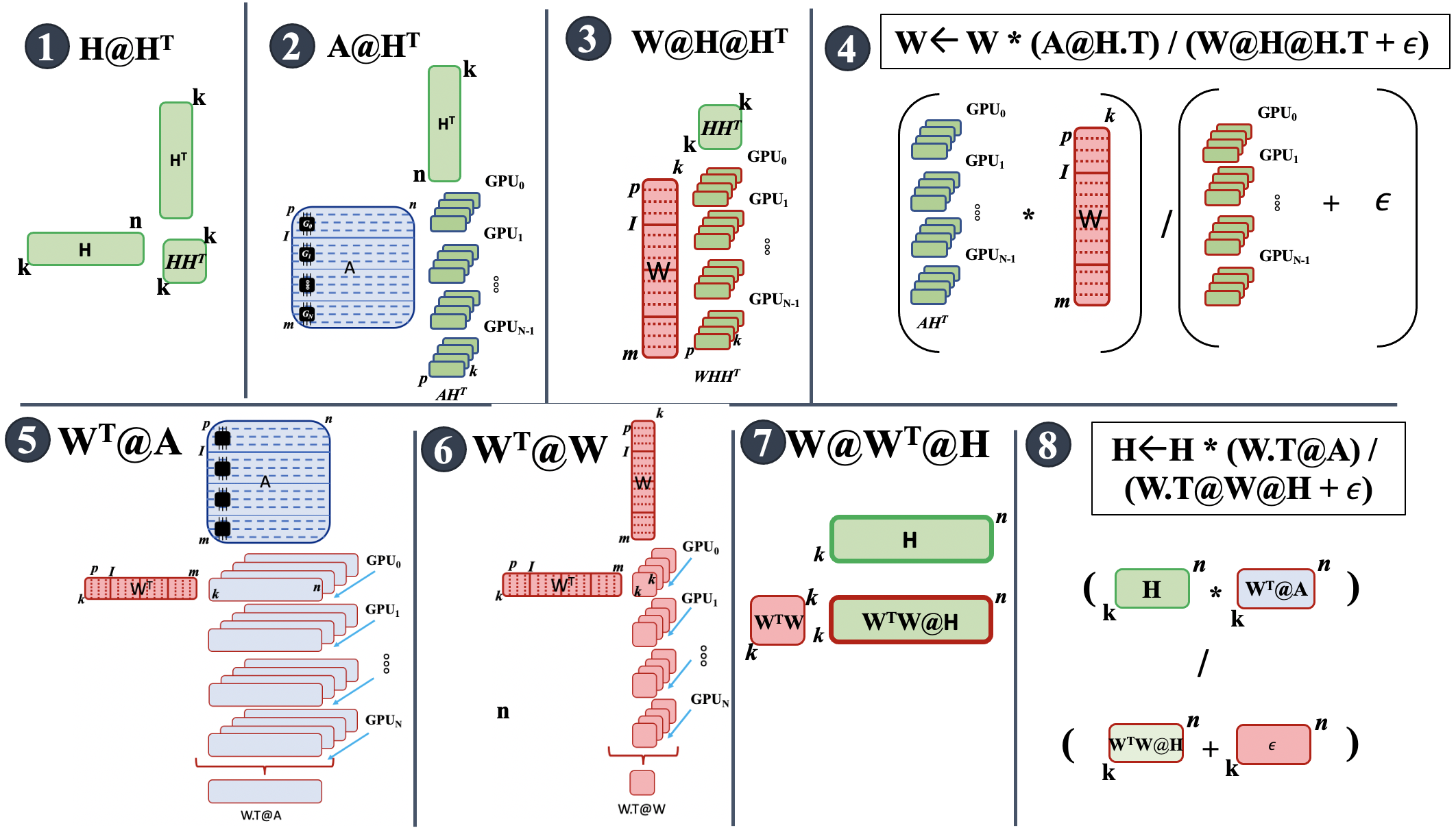}} 
\caption{Illustration of the batched multiplicative update Algorithm~\ref{alg:batched_RNMF}  for the row partition(RNMF) and colinear batching. Green array is duplicated across different MPI ranks. Blue and red arrays are distributed, and only red array is cached on device. For RNMF, p is Out-of-Memory batch width and J is distributed partition width.}
\label{fig:MU_batch_row_algo2}

\end{figure*}

\begin{algorithm}[!t]
\scriptsize
    \caption{$\mat{W},\mat{H}= \operatorname{CNMF}(\mat{X},k)$--\emph{CNMF} with \emph{orthogonal} batching}
    \label{alg:batched_OCNMF}
	\begin{algorithmic}[1]
    \Require  $\mat{X} \in \mathbb{R}_{+}^{m \times n}$, $k$ is the rank of approximation and $max\_iters$ is the number of iterations.
    \Require $\mat{X}$ distributed across $N$ GPUs where  $\mat{X}_{j} \in \mathbb{R}^{m \times n/N}$ where $J=\frac{n}{N}$ if $m<n$. Locally to each GPU, $\mat{X}_{j}$ can be split into $n_b$ batches where $n_b=\frac{m}{p}$ following a \emph{orthogonal} batching strategy where the batch is of shape $p \times J$. Similarly the co-factor $\mat{W}$ local to each GPU given by $\mat{W}\in \mathbb{R}^{m \times k}$ is locally divided into $n_b$ batches where the batch is of shape $p \times k$. $\mat{H}$ is distributed across N GPUS such that $\mat{H}_j \in \mathbb{R}^{k \times J}$. 

    \State Initialize $\mat{W}_i$,$\mat{H}_j$ = $\operatorname{rand}(I,k)$,$\operatorname{rand}(k,J)$ 
    \State Initialize \textbf{SQ}, a queue of CUDA-streams of size $q_s$.
    \State Initalize zero array accumulators $\ten{WTA}\in \mathbb{Z}_{}^{q_s \times k\times n}$ and  $\ten{WTW}\in \mathbb{Z}_{}^{q_s \times k\times k} $
    \For{l in [1,$max\_iters$]} \label{labels:batched:CNMF:iter_loop}
     \Statex \textbf{/* Update  $\mat{H}$ given $\mat{W}$ */}

    \For {$b$ in  $n_b$} 

         \State \textbf{SQ} $->$ \textit{st} \Comment{De-queue stream st from \textbf{SQ}} \label{labels:batched:CNMF:de_Q}
           \While{in context \textit{st}} \Comment{Calculations in loop are in non-default stream st} \label{labels:batched:CNMF:stm_context}
             \Statex $i_0=b*p,\ i_1 = (b+1)*p$, $j_0=gID*J,\ j_1 = (gID+1)*J$ 
             \State $\mat{A}$ = $\operatorname{H2D}(\mat{X}_j[i0:i1,:], \textit{st})$
             \Comment{ $\operatorname{H2D}(\mat{x},\textit{st})$ stands for async copy of $\mat{x}$ from Host to GPU using non-default stream \textit{st}} \label{labels:batched:CNMF:H2D_A_1} 
               
             \State $\mat{W}_b$ = $\operatorname{H2D}(\mat[l][]{W}[i0:i1, :], \textit{st})$  \label{labels:batched:CNMF:H2D_W_1}
             \State $\ten{WTA}[st]$ += $\mat{W}_{b}^T@\mat{A}$ \Comment{Accumulate local $\mat{W}^T@\mat{A}$} \label{labels:batched:CNMF:accum_WTA}
             \State $\ten{WTW}[st]$ += $\mat{W}_{b}^T \mat{W}_b$ \Comment{Accumulate local $\mat{W}^T@\mat{W}_b$} \label{labels:batched:CNMF:accum_WTW}
             \State  \textit{st} $->$ \textbf{SQ}  \Comment{En-queue stream st back into \textbf{SQ}, exit context st} \label{labels:batched:CNMF:en_Q}    
        \EndWhile
     \EndFor
  
    \State $\mat{WTA}$ = $\sum_{i=1}^{n_b} \ten{WTA}_{i,:,:}$\Comment{ Reduce of $\ten{WTA}$ local to each GPU} \label{labels:batched:CNMF:reduce_WTA} 
    \State $\mat{WTW}$ = $\sum_{i=1}^{n_b} \ten{WTW}_{i,:,:}$ \Comment{ Reduce of $\ten{WTW}$ local to each GPU} \label{labels:batched:CNMF:reduce_WTW} 

    \State $\mat[l+1][j]{H} \times = \mat{WTA} / (\mat{WTW} @ \mat[l][j]{H} + \epsilon) $ \ \ \ \ \ \ \ \ \ \ \ \ 
         \Statex \textbf{/* Update  $\mat{W}$ given $\mat{W}$ */}
    \State $\mat{HHT} = \mat[l+1][j]{H} @\mat{H^{T^{(l+1)}}}_j$                            
    \State $\mat{HHT} \leftarrow \operatorname{All\_Reduce\_sum}(\mat{HHT})$
    \For {$b$ in $0$ to $n_B$} 
             \State \textbf{SQ} $->$ \textit{st} \Comment{De-queue stream st from \textbf{SQ}}
           \While{in context \textit{st}} \Comment{Calculations in loop are in non-default stream st}
 
        \State Set $i_0=b*p,\ i_1 = (b+1)*p$
       \State $\mat{A}$ = $\operatorname{H2D}(\mat{X}_j[i0:i1, :], \textit{st})$
             \Comment{ $\operatorname{H2D}(\mat{x},\textit{st})$ stands for async copy of $\mat{x}$ from Host to GPU using non-default stream \textit{st}}   \label{labels:batched:CNMF:H2D_A_2}
        \State $\mat{W}_b$ = $\operatorname{H2D}(\mat[l][]{W}[i0:i1, :], \textit{st})$  \label{labels:batched:CNMF:H2D_W_2}
       \State $\mat{WHHT}_b$ = $\mat{W}_b@\mat{HHT} + \epsilon$
        \State $\mat{AHT}$ = $\mat{A}@\mat{H^{T^{(l+1)}}}_j$
        \State $\mat{AHT} =\operatorname{All\_Reduce\_sum}(\mat{AHT},stream=st)$ \Comment{Perform stream aware Reduce with NCCL} \label{labels:batched:CNMF:ar_AHT} \ \ \ \ \ \ \ \ \,
        \State $\mat{W}_b \times = \mat{AHT}/(\mat{WHHT}_b+\epsilon)$ \ \ \ \ \ \ \
        \State $\mat[l+1][]{W}_i[i0:i1, :]$ = $\operatorname{D2H}(\mat{W}_b, st)$         \Comment{ $\operatorname{D2H}(\mat{x},\textit{st})$ stands for async copy of $\mat{x}$ from GPU to Host using non-default stream \textit{st}.} 
        \EndWhile
     \EndFor
     \EndFor
\end{algorithmic} 

\end{algorithm}
\subsection{Out-of-memory implementation and memory complexity analysis}
\label{subsec:dist_batching}

In \emph{pyDNMF-GPU}, \emph{OOM-0} problems are handled using a tiling approach where temporary results like $\mat{A}\mat{H}^T$, $\mat{W}^T\mat{A}$ or $\mat{W}\mat{H}$ are evaluated in small chunks, by tiling one of the operands, such that the size of the tile sets the memory required for the calculation. In RNMF for instance, the criterion $\Arrowvert \mat{A} - \mat{W} \mat{H} \Arrowvert_F \leq \eta$, can be evaluated in $m/p$ small chunks obtained by tiling $\mat{W}$ into smaller tiles of size $p \times k$. This results in computing $nt$ chunks of $[\Arrowvert \mat{A} - \mat{W} \mat{H} \Arrowvert_F]_p$ which are accumulated into the total error $e$ such that $e=\sum{([\Arrowvert \mat{A} - \mat{W} \mat{H} \Arrowvert_F]_t)}_{t=0}^{m/p-1}$, which can later be used to check the conversion condition $e \leq \eta$. This allows the reduction of the memory required to check the conversion criterion from $\mathcal{O}(m \times n)$ to $\mathcal{O}(p \times n)$. Because all matrices involved in the calculations are stored on GPU memory, performance loss due to tiling can be negligible, especially on modern GPU architecture like NVIDIA Ampere A100, which uses low latency and high bandwidth HBM memory. Using the tiling approach, the memory required to perform NMF on GPU can be reduced from $S_{NMF} \sim 4 S_A$ to approximately $2 S_A \leq S_{NMF} \leq 3 S_A$.

\begin{algorithm}[t!]
\scriptsize
    \caption{$\mat{W},\mat{H}= \operatorname{RNMF}(\mat{X},k)$--\emph{RNMF} with \emph{co-linear} batching}
    \label{alg:batched_RNMF}
	\begin{algorithmic}[1]
    \Require  $\mat{X} \in \mathbb{R}_{+}^{m \times n}$, $k$ is the rank of approximation and $max\_iters$ is the number of iterations.
    \Require $\mat{X}$ distributed across $N$ GPUs where  $\mat{X}_{i} \in \mathbb{R}^{I \times n}$ where $I=\frac{m}{N}$ if $m>n$. Locally to each GPU, $\mat{X}_{i}$ can be split into $n_b$ batches following a \emph{co-linear} batching strategy where the batch size is $b_s = \frac{I}{n_b}\times n$. Similarly the co-factor $\mat{W}$ local to each GPU given by $\mat{W}_{i}\in \mathbb{R}^{I \times k}$ is locally divided into $n_b$ batches where $b_s = \frac{I}{n_b}\times k$. $\mat{H}\in \mathbb{R}^{k \times n}$ is cached and replicated across the GPUs. 

    \State Initialize $\mat{W}_i$,$\mat{H}$ = $\operatorname{rand}(I,k)$,$\operatorname{rand}(k,n)$ 
    \State Initialize \textbf{SQ}, a queue of CUDA-streams of size $q_s$.
    \State Initalize accumulators $\ten{WTA}\in \mathbb{R}_{+}^{q_s \times k\times n}$ and  $\ten{WTW}\in \mathbb{R}_{+}^{q_s \times k\times k} $
    \For{l in [1,$max\_iters$]}
     \Statex \textbf{/* Update  $\mat{W}$ given $\mat{H}$ */}
    \State $\mat{HHT}$ = $\mat[l][]{H}@\mat{H^{T^{(l)}}}$
    \For {$p$ in $n_b$}
       \State $io,i_1$ = $p*b_s,(p+1)*b_s$
       \State \textbf{SQ} $->$ \textit{st} \Comment{De-queue stream st from \textbf{SQ}}
       \While{in context \textit{st}} \Comment{Calculations in loop are in non-default stream st}
         \State $\mat{A}$ = $\operatorname{H2D}(\mat{X}_{i}[i0:i1, :], \textit{st})$
         \Comment{ $\operatorname{H2D}(\mat{x},\textit{st})$ stands for async copy of $\mat{x}$ from Host to GPU using non-default stream \textit{st}}
         \State $\mat{W}_b$ = $\operatorname{H2D}(\mat[l][]{W}_i[i0:i1, :], \textit{st})$ 
         \State $\mat{AHT}$ = $\mat{A}@\mat{H^{T^{(l)}}}$
         \State $\mat{WHHT}$ = $\mat{W}_b@\mat{HHT} + \epsilon$
         \State $\mat{W}_b \times = \mat{AHT}/\mat{AWWT}$\Comment{$\mat{W}$ update}
         \State $\mat[l+1][]{W}_i[i0:i1, :]$ = $\operatorname{D2H}(\mat{W}_b, st)$  
         \Comment{ $\operatorname{D2H}(\mat{x},\textit{st})$ stands for async copy of $\mat{x}$ from GPU to Host using non-default stream \textit{st}.}
         \State $\ten{WTA}[st]$ += $\mat{W}_{b}^T@\mat{A}$ \Comment{Accumulate local $\mat{W}^T@\mat{A}$} \label{labels:batched:RNMF:accum_WTA}
         \State $\ten{WTW}[st]$ += $\mat{W}_{b}^T \mat{W}_b$ \Comment{Accumulate local $\mat{W}^T@\mat{W}_b$} \label{labels:batched:RNMF:accum_WTW}
         \State  \textit{st} $->$ \textbf{SQ}  \Comment{En-queue stream st back into SQ, exit context st}
       \EndWhile
    \EndFor
     \Statex \textbf{/* Update  $\mat{H}$ given $\mat{W}$ */}
    \State $\mat{WTA}$ = $\sum_{i=1}^{n_b} \ten{WTA}_{i,:,:}$\Comment{ Reduce of $\ten{WTA}$ local to each GPU} \label{labels:batched:RNMF:local_red_WTA} 
    \State $\mat{WTW}$ = $\sum_{i=1}^{n_b} \ten{WTW}_{i,:,:}$ \Comment{ Reduce of $\ten{WTW}$ local to each GPU} \label{labels:batched:RNMF:local_red_WTW} 
    \State $\mat{WTA}$ = $\operatorname{All\_Reduce\_sum}(\mat{WTA})$ \Comment{Global Reduce of $\mat{WTA}$ across all GPUs} \label{labels:batched:RNMF:global_red_WTA}
    \State $\mat{WTW}$ = $\operatorname{All\_Reduce\_sum}(\mat{WTW})$ \Comment{Global Reduce of $\mat{WTW}$ across all GPUs} \label{labels:batched:RNMF:global_red_WTW}
    \State $\mat{WTWH} = \mat{WTW}@\mat[l][]{H}$
    \State $\mat{WTWH} = \mat{WTW}@\mat[l][]{H} + \epsilon$
    \State $\mat[l+1][]{H}  \times =  \mat{WTA}/\mat{WTWH}$ \Comment{$\mat{H}$ update}
\EndFor
\end{algorithmic} 

\end{algorithm}

When dealing with \emph{OOM-1} cases, light arrays are cached on GPU memory, and heavier arrays are kept on host memory and batched to respective GPUs as needed. Further, an appropriate batching strategy for the chosen memory partition is required to limit unnecessary \emph{D2H} and \emph{H2D} copies. In \emph{PyDNMFk-GPU}, we employ a \emph{1D} \emph{co-linear} batching strategy, illustrated in Figure~\ref{fig:nmf_row_part}, where the elements in the batch are arrays of length equal $max(m,n)$. This batching strategy turns out to employ half the \emph{D2H} and \emph{H2D} memory copies required by an \emph{orthogonal} batching strategy, illustrated in Figure~\ref{fig:nmf_col_part} for the column partition, where the elements in the batch are vectors of length equal $min(m,n)$. Let $p$ be a batch size control parameter. In \emph{RNMF} (\emph{CNMF}) the number of batches is then given by $n_B = m/p$ ($n_B = n/p$). In the extreme case where both $m$ and $n$ are very large, only the light array, $\mat{W}[J, :]$ is cached on GPU memory, and heavier arrays $\mat{A}[J, b_0:b_i]$ ($A[b_0:b_1, J]$) and $\mat{H}[:, b_0:b_1]$ ($\mat{W}[b_0:b_1, :]$) batched to their respective GPUs, such that for the $b^{th}$ batch, $b_0=b \times p$ and  $b_1=(b+1) \times p$. 

An implementation of the distributed \emph{CNMF} with \emph{orthogonal} batching is given in Algorithm~\ref{alg:batched_OCNMF}. The calculation of the different intermediate products is illustrated in Figure~\ref{fig:MU_batch_col_algo}, where batch delimitation is represented with dashed lines. The top row shows all intermediate products computed during $\mat{H}$-update, and products computed in $\mat{W}$-update is shown in the bottom row. Intermediate products $\mat{W}^T @ \mat{A}$ and $\mat{W}@\mat{W}^T$ can be computed with $n_B$ independent batches each containing $[\mat{W}^T @ \mat{A}]_b$ and $[W^T @ W]_b$ sub-products. Each batch is queued to a non-default CUDA stream $Stm_b$ along with the transfer of $\mat{A}_b[b_0:b_i,J]$ and $\mat{W}_b[b_0:b_i, :]$, and when calculated, each sub-product is added to a local accumulator (see lines~\ref{labels:batched:CNMF:accum_WTA}-\ref{labels:batched:CNMF:accum_WTW} of Algorithm~\ref{alg:batched_OCNMF}). Once all batches have been processed, all accumulators are reduced to obtain the full values of $W^TW$ and $W^TA$, (see lines~\ref{labels:batched:CNMF:reduce_WTA}-\ref{labels:batched:CNMF:reduce_WTW} of Algorithm~\ref{alg:batched_OCNMF}). Note that this reduction is local to each GPU and does not involve communication. Special batch en-queuing and de-queuing policies are implemented with CUDA events, so as to limit (control) the number of concurrent batches on GPU to $q_s$ (see lines~\ref{labels:batched:CNMF:de_Q}-\ref{labels:batched:CNMF:stm_context},\ref{labels:batched:CNMF:en_Q} of Algorithm~\ref{alg:batched_OCNMF}). This way, the memory requirement for $H_{update}$ is bounded by $q_s \times [p\times J]$, as $\mat{W}^T\mat{W}@\mat{H}$ and $\mat{H}*(\mat{W}^T\mat{A})/(\mat{W}^T\mat{W}\mat{H} + \epsilon)$ have a $k \times J$ memory requirement. This is important, especially when dealing with large sparse arrays, which can be cheap to cache on the device but can also have co-factors becoming prohibitively expensive to cache when $k$ becomes large. For instance, in \emph{CNMF}, when $m \sim 10 \ million$, the size of $\mat{H}$ will approximate $20GB$ in single precision when $k \sim 512$ .

Intermediate products $\mat{A}@\mat{H}^T$ and $\mat{W}@\mat{H}\mat{H}^T$ of the $\mat{W}$-update are computed similarly to $\mat{W}^T @ \mat{A}$ and $\mat{W}@\mat{W}^T$, except $\mat{A}@\mat{H}^T$ will require an intermediate $all-reduce-sum$ of sub-products $[\mat{A} @ \mat{H}^T]_b$ of batches of same stream number from the different GPUs (see line~\ref{labels:batched:CNMF:ar_AHT} of Algorithm~\ref{alg:batched_OCNMF}). The resulting memory complexity of this implementation is found to be of the order of $\mathcal{O}(p \times n \times q_s)$ when $p >> k$ which is the aggregated memory utilization caused by the $q_s$ concurrent uploads of batches of $\mat{A}$ of size $p \times n$ at line~\ref{labels:batched:CNMF:H2D_A_1} or line~\ref{labels:batched:CNMF:H2D_A_2} of Algorithm~\ref{alg:batched_OCNMF}. This is a significant saving compared to the estimated $S_{NMF} \sim 3 \times S_A$ when not checking the convergence condition $\Arrowvert \mat{A} - \mat{W} \mat{H} \Arrowvert_F \leq \eta$. When the convergence criterion is checked, the error computation is tiled similarly as it was done for \emph{OOM-0} scenarios, resulting in a memory utilisation $S_{NMF} \sim 2 \times p \times n \times q_s$ when $p >> k$.

Note that the use of batches here will only increase $intra-node$ communication due to mem-copies, as it is not possible to cache $\mat{A}$ and $\mat{W}$ on the device, however major shortcomings of using the \emph{orthogonal} batching can be pointed out through the example of Algorithm~\ref{alg:batched_OCNMF} discussed above. First, the need to upload batches two times at lines (\ref{labels:batched:CNMF:H2D_A_1}-\ref{labels:batched:CNMF:H2D_W_1} and lines~\ref{labels:batched:CNMF:H2D_A_2}-\ref{labels:batched:CNMF:H2D_W_2} of Algorithm~\ref{alg:batched_OCNMF}) is very inefficient as the second set of H2D will significantly (almost double) data transfer costs. Second, unnecessary additional latency due to load balancing delays when the streams are scheduled in a different order on the different GPUs can occur at line~\ref{labels:batched:CNMF:ar_AHT} of Algorithm~\ref{alg:batched_OCNMF}. Above all, the worst result here is that both inefficiencies multiply with the number of iterations ( see line~\ref{labels:batched:CNMF:iter_loop} of Algorithm~\ref{alg:batched_OCNMF}).

A better implementation uses a \emph{co-linear} batching strategy as it is done in the batched implementation of the distributed \emph{RNMF} given in Algorithm~\ref{alg:batched_RNMF}. The calculation of the different intermediate products is illustrated in Figure~\ref{fig:MU_batch_row_algo2}. The top row shows all intermediate products computed during $\mat{W}$-update, and products computed in $\mat{H}$-update is shown in the bottom row. The $\mat{W}$-update (cartoons 1-4 of Figure~\ref{fig:MU_batch_row_algo2} is embarrassingly parallel and can be done at a batch level. This means that within each batch, we have the updated partition of $\mat{W}$ readily available to compute local sub-products $\mat{W}^T@\mat{A}$ and $\mat{W}^T@\mat{A}$ in the $\mat{H}$-update. This avoids the need for a second data upload, as was the case with implementation using an \emph{orthogonal} batching strategy. Further, the aggregation of $\mat{W}^T@\mat{A}$ and $\mat{W}^T@\mat{A}$ first consists of a local accumulation of the sub-products (lines~\ref{labels:batched:RNMF:accum_WTA}-\ref{labels:batched:RNMF:accum_WTW} of Algorithm~\ref{alg:batched_RNMF}) followed by a local reduction (lines~\ref{labels:batched:RNMF:local_red_WTA}-\ref{labels:batched:RNMF:local_red_WTW} of Algorithm~\ref{alg:batched_RNMF}), then a global reduction (lines~\ref{labels:batched:RNMF:global_red_WTA}-\ref{labels:batched:RNMF:global_red_WTW} of Algorithm~\ref{alg:batched_RNMF}) illustrated in cartoons 5-6 of Figure~\ref{fig:MU_batch_row_algo2}.  This does not require communication between batches of the same stream number and consequently avoids load balancing issues as discussed above in case using an \emph{orthogonal} batching strategy.

\section{Benchmarks results and discussion}
\label{sec:results}
\subsection{ Hardware infrastructure and software environment}
Benchmark tests were performed on three different HPC clusters to illustrate the portability and scalability of \textit{pyDNMF-GPU}. 
The first cluster, \emph{Kodiak}, is a  LANL internal HPC cluster with 133 compute nodes with dual Xeon E5-2695 v4 CPUs and four NVIDIA Pascal P100 GPGPUs each. Each NVIDIA Pascal P100 GPGPU has 16GB VRAM and uses PCI-E 16X gen 3 Links. The cluster peaks at  1850TF/s and uses an Infiniband interconnect. Each GPU peaks at 9.3 teraflops for single precision.
The second cluster, \emph{Chicoma}, is also a LANL internal HPC cluster, composed of 118 compute nodes where each node has  2 AMD EPYC 7713 Processors and 4 NVIDIA Ampere A100 GPUs. The AMD EPYC 7713 CPUs have 64 cores peaking at 3.67 GHz and 256 GB RAM. Each of the four NVIDIA A100 GPUs in each node provides a theoretical double-precision arithmetic capability of approximately 19.5 teraflops with 40GB VRAM memory. The nodes are networked with HPE/Cray slingshot 10 interconnect with 100Gbit/s bandwidth. \emph{Chicoma} runs Shasta 1.4 OS and SLURM Job manager. The third cluster, \emph{Summit}, peaks at over 200 petaflops in double-precision theoretical performance and comprises 4600 IBM AC922 compute nodes, with two IBM POWER9 CPUs and six NVIDIA Volta V100 GPUs each which peak at 15.7 single precision. The POWER9 CPUs have 22 cores running at 3.07 GHz. The six NVIDIA Tesla V100 GPUs in each node provide a theoretical double-precision arithmetic capability of approximately 40 teraflops with VRAM memory of 16GB/GPU. Dual NVLink 2.0 connections between CPUs and GPUs provide a 25-GB/s transfer rate in each direction on each NVLink, yielding an aggregate
bidirectional bandwidth of 100 GB/s. The nodes are networked in a non-blocking fat-tree topology by Infiniband. Summit deploys an RHEL 7.4 OS and IBM Job step manager jsrun to run compute jobs. Jsrun provides a fine control of how node-level resources are allocated on these systems, including CPU cores, GPUs, and hardware threads.

\textit{pyDNMF-GPU} is written in python and uses other off the shelf python libraries such as CuPy~\cite{cupy_learningsys2017}, Numpy~\cite{harris2020array},  MPI4PY~\cite{MPI4PY} and Scipy~\cite{2020SciPy-NMeth}. It supports dense and sparse datasets on various hardware architectures and handles communication using a low-latency NCCL-based communicator. NCCL is an open-source library providing inter-GPU communication primitives developed and maintained by NVIDIA. NCCL performs automatic hardware topology detection, which it then uses in graph search algorithms to identify communication paths that offer the highest bandwidth and lowest latencies for communication between GPUs intra- and inter-node (e.g., between GPUs that are on the same compute node, as well as between GPUs that are on separate compute nodes). NCCL is compatible with many multi-GPU parallelization models, and provides the ability to perform MPI-like collective and point-to-point operations such as allgather, reduce, broadcast, allreduce, send, and recv. NCCL was initially proposed to help with the need to transfer large message GPU buffers in deep learning applications efficiently. Many leading deep learning frameworks like Chainer, PyTorch, and TensorFlow have since integrated NCCL to accelerate deep learning training on multi-GPU, and multi-node systems, which has motivated us to use NCCL to handle communication in our work. All implementations discussed in the section above were found to benefit from a reduction in data transfer latency and communication performance (both intra-node and inter-node communications), using our low latency NCCL-based communicators versus MPI. An example of such benefit in communication performance gain is illustrated in the subsection~\ref{subsec:cpu_vs_gpu} below that compares the new NMF implementation proposed in this work that uses an NCCL-based communicator to the prior \emph{pyDNMFk} that uses a traditional MPI based communicator. A More comprehensive and detailed comparative study between NCCL and MPI can be found in the analysis by Awan \cite{awan2016efficient}.

\label{sec:Results}
\subsection{Performance benchmark results of \emph{pyDNMF-GPU} vs \emph{pyDNMFk}}
\label{subsec:cpu_vs_gpu}
\begin{figure}[t!]
    \centering
    \begin{minipage}[b]{0.48\linewidth}

        \includegraphics[width=1\linewidth]{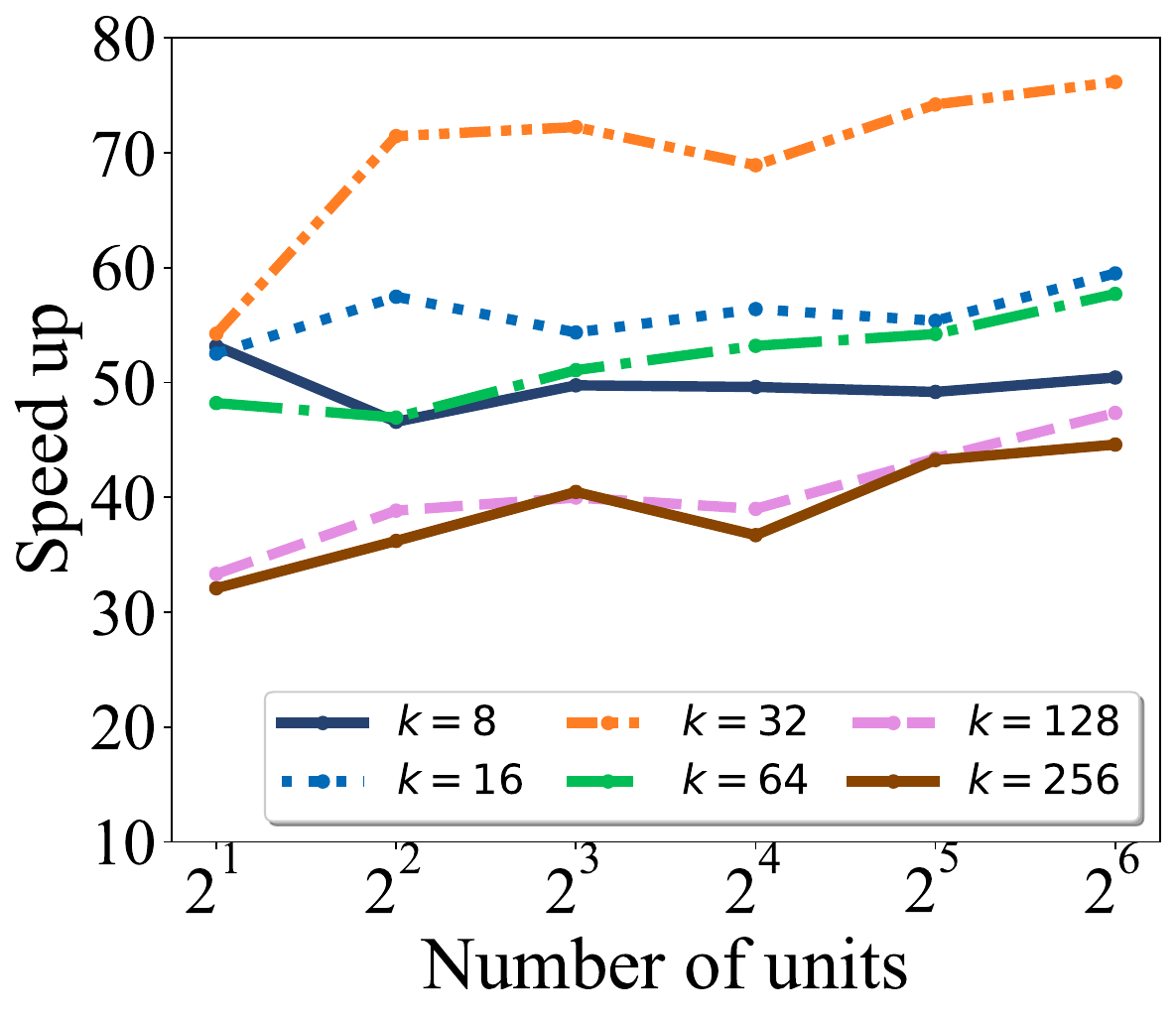}
        \subcaption{Speedups of the compute times for various latent dimensions on different numbers of compute units. \label{fig:cpu_vs_gpu_nmf}}
    \end{minipage}
    \hspace{1mm}
    \begin{minipage}[b]{0.48\linewidth}

        \includegraphics[width=1\linewidth]{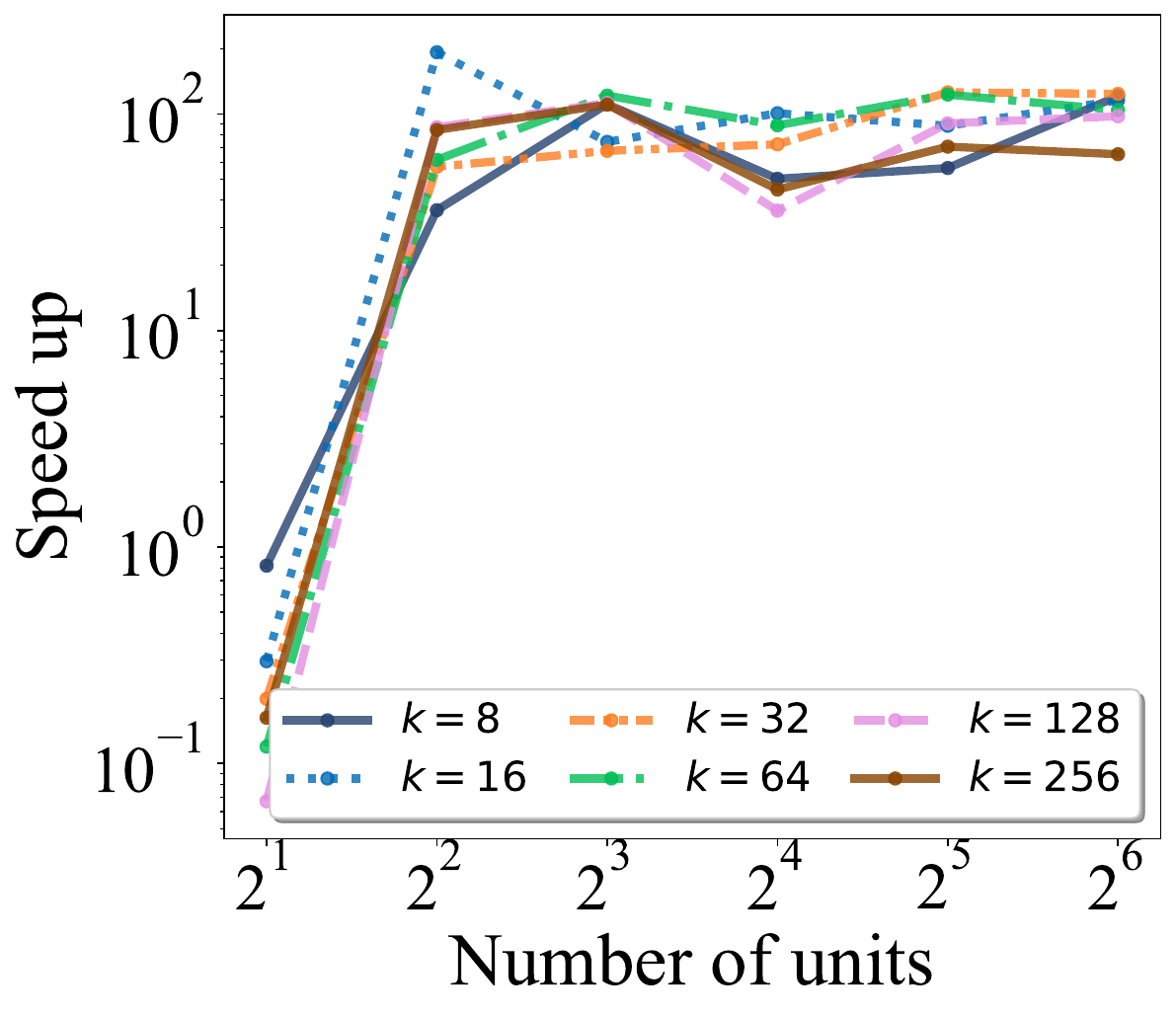}
        \subcaption{Speedups of the communication times for various latent dimensions on different numbers of compute units. \label{fig:cpu_vs_gpu_comm}}
    \end{minipage}

    \caption{ 
    Results of benchmarking experiment showing speedup gain using N GPUs vs N CPUs, for various $k$. Speedup gained on NMF calculation time is shown in Figure~\ref{fig:cpu_vs_gpu_nmf} and speedup gained on communication time is shown in Figure~\ref{fig:cpu_vs_gpu_comm}.
    }
    \label{fig:cpu_vs_gpu}
    \vspace{-1em}
\end{figure}

The performance gained using GPU over CPU is assessed with speedup computed as the ratio of time measured on CPU with \textit{pyDNMFk}\cite{bhattarai2021pydnmfk}, to time measured on GPU with \textit{pyDNMF-GPU}. For this study, we used a dense matrix of shape and size $S_A$ of memory (in bytes) that respectively scale as $[N \times 65536, 32768]$ and  $N \times 8GB$, where $N$ is the number of GPU or CPU units. Speedup measured on the \emph{Kodiak} cluster are reported in Figure~\ref{fig:cpu_vs_gpu}.  Figure~\ref{fig:cpu_vs_gpu_nmf} shows speedup in NMF time as a function of the number of units for various $k$ . First, we note an increasing speedup with the increasing number of units, and second, we note a decreasing performance with increasing $k$ when $k \geq 32$. The low performance observed at $k < 32$ is explained by low GPU occupancy. The best performance is obtained when $k=32$, peaking at $~76X$.
We also report speedup in communication time computed as the ratio of total communication time measured with \textit{pyDNMFk} to the total communication time measured \textit{pyDNMF-GPU}. The former used MPI based communicator and the latter used an NCCL-based communicator. Speedup in communication time is reported as a function of number of units for various $k$ in Figure~\ref{fig:cpu_vs_gpu_comm}. We note $\sim 80X-100X$ speedup when $N>2$, the number of units above which inter-node communications start. This clearly shows a significant performance gain in communication when using NCCL in \textit{pyDNMF-GPU} over MPI in \textit{pyDNMFk}.
\subsection{Strong and Weak scalability of \emph{pyDNMF-GPU}}

The scalability of the proposed NMFk algorithm is assessed using both strong and weak scaling analysis. This scaling study measures NMF execution time for a given problem size as a function of the number of compute units. 
Compute nodes (with 4 GPUs each) are chosen as compute units in strong scaling analysis, while individual GPUs are chosen as compute units in weak scaling analysis. The problem size $S_A$ is chosen to use most of the available 16GB VRAM per GPU. To this end, $S_A$ is fixed at $S_A \approx 4\times 8GB = 32GB$ in strong scaling analysis and chosen to scale as $S_A \approx 8GB \times N$ in weak scaling analysis. This is accomplished by generating a random synthetic array $A$ of shape $[4 \times 65536, 32768]$ and $[N \times 65536, 32768]$ respectively in both strong and weak scaling. Cases of sparse $A$ with density $10^{-5}$ were also studied, and for those cases, $A$ was generated as a random synthetic array of shape $[4 \times 2097152, 65536]$ in strong scaling analysis, and of shape $[N \times 2097152, 65536]$ was chosen in weak scaling analysis.

\subsubsection{Strong scalability}
\begin{figure}[t!]
\centering
    \begin{minipage}[b]{0.48\linewidth}
        \includegraphics[width=1\linewidth]{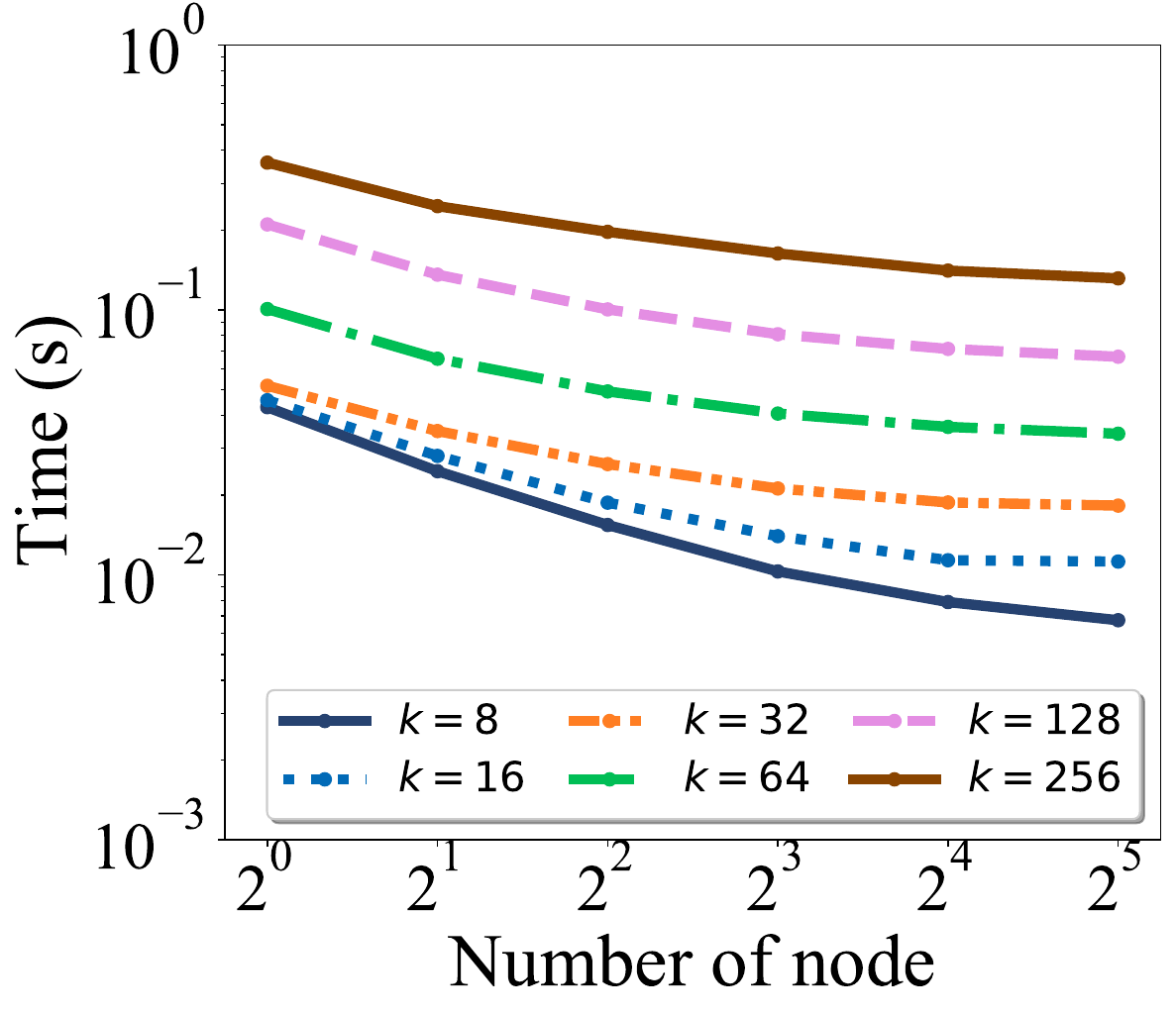}
        \subcaption{Dense \label{fig:strong_dense_n_t_k}}
    \end{minipage}
    \hspace{1mm}
    \begin{minipage}[b]{0.48\linewidth}
        \includegraphics[width=1\linewidth]{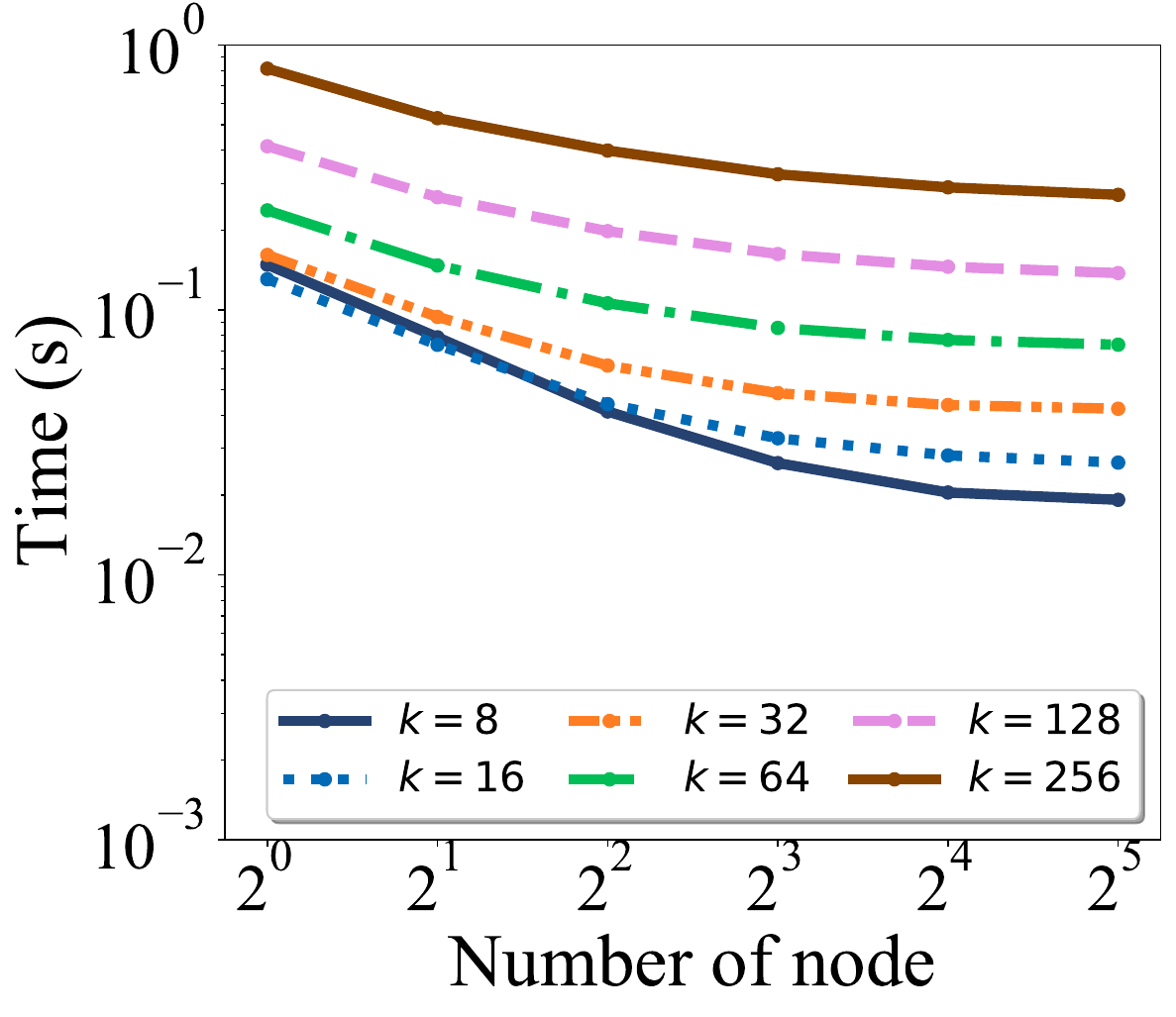}
        \subcaption{Sparse \label{fig:strong_sparse_n_t_k}}
    \end{minipage}
    \hspace{1mm}
    \begin{minipage}[b]{0.48\linewidth}
        \includegraphics[width=1\linewidth]{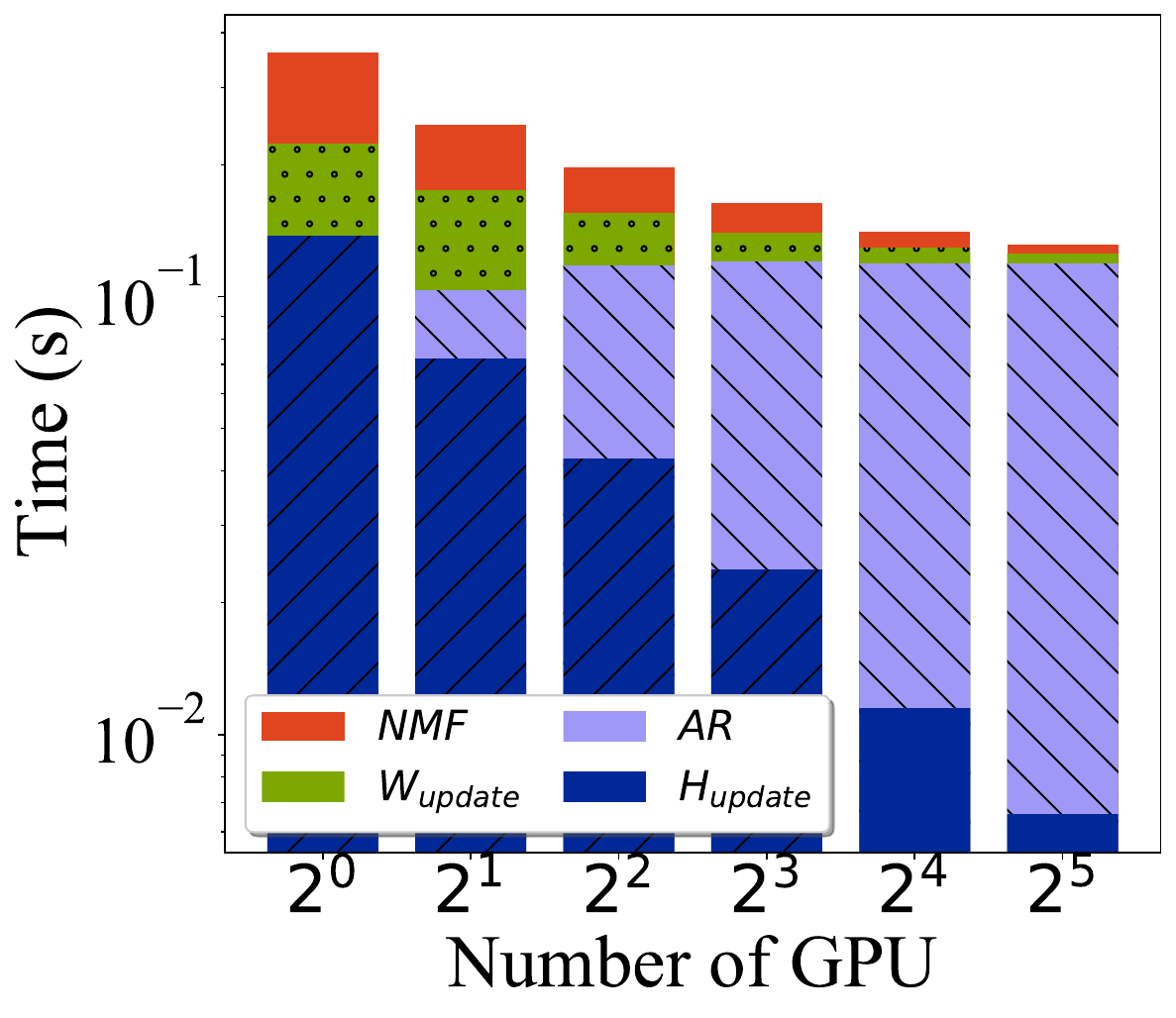}
        \subcaption{Dense \label{fig:strong_dense_n_t}}
    \end{minipage}
    \hspace{1mm}
    \begin{minipage}[b]{0.48\linewidth}
        \includegraphics[width=1\linewidth]{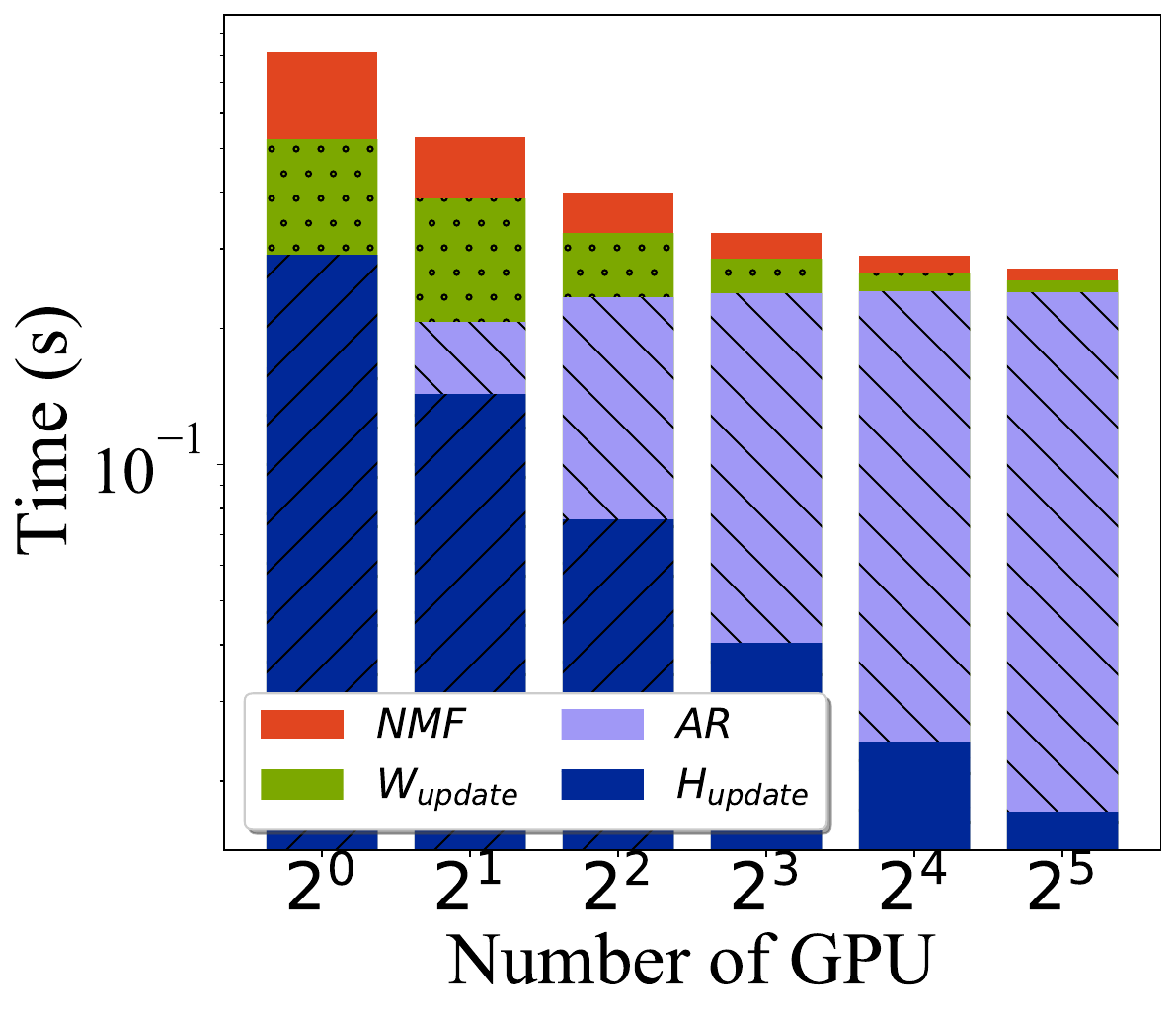}
        \subcaption{Sparse \label{fig:strong_sparse_n_t}}
    \end{minipage}
    \caption{ Results of strong scaling study performed on \emph{Kodiak}. NMF time vs number of node for various $k$ dense and sparse $\mat{A}$ respectively shown in (\ref{fig:strong_dense_n_t_k}) and (\ref{fig:strong_sparse_n_t_k}). For the case $k=8$, execution time of $H_{update}$, $W_{update}$ and All-reduce communication are compared in (\ref{fig:strong_dense_n_t}) and (\ref{fig:strong_sparse_n_t}), respectively for dense and sparse $\mat{A}$. \label{fig:strong_scaling}}
    \vspace{-1em}
\end{figure}

Strong scaling results for cases where $k=8,16,32,64,128,256$ are shown Figure~\ref{fig:strong_dense_n_t_k}. NMF time is found to increase with $k$ and to decrease with the increasing number of compute nodes. Good strong scaling is indicated by a linear decrease of NMF time with increasing compute grid size, and such behavior is only observed in select parts of the obtained results. Strong scaling is maintained up to a count of 8 nodes when $k=8$, then to 4 nodes when $k=16$, and lost when $k>16$. Identical scaling is observed for cases where $A$ is sparse, as shown in Figure~\ref{fig:strong_sparse_n_t_k}.
 The worst case scenarios, when $k=256$, can be diagnosed from breakdown of $H_{update}$, $W_{update}$ and combined all-reduce-sum ($AR$) execution time, as detailed in Figure~\ref{fig:strong_dense_n_t}. $H_{update}$ is shown to maintain good scaling at all compute grid sizes, while $W_{update}$ had poor scaling at each tested compute grid size. $W_{update}$'s poor scaling is strongly influenced by $AR$ communications time, which already makes up more than $80\%$ of $W_{update}$ at 2 node count, which increases non-linearly with node count. At full grid size, $AR$ time makes up more than $98\%$ of  $W_{update}$, influencing the overall NMF time dominated by $W_{update}$ time. The same explanation applies to cases where $A$ is sparse, as one can interpret from Figure~\ref{fig:strong_sparse_n_t}.

\subsubsection{Weak scalability}
\begin{figure}[t!]
\centering
    \begin{minipage}[b]{0.48\linewidth}
        \includegraphics[width=1\linewidth]{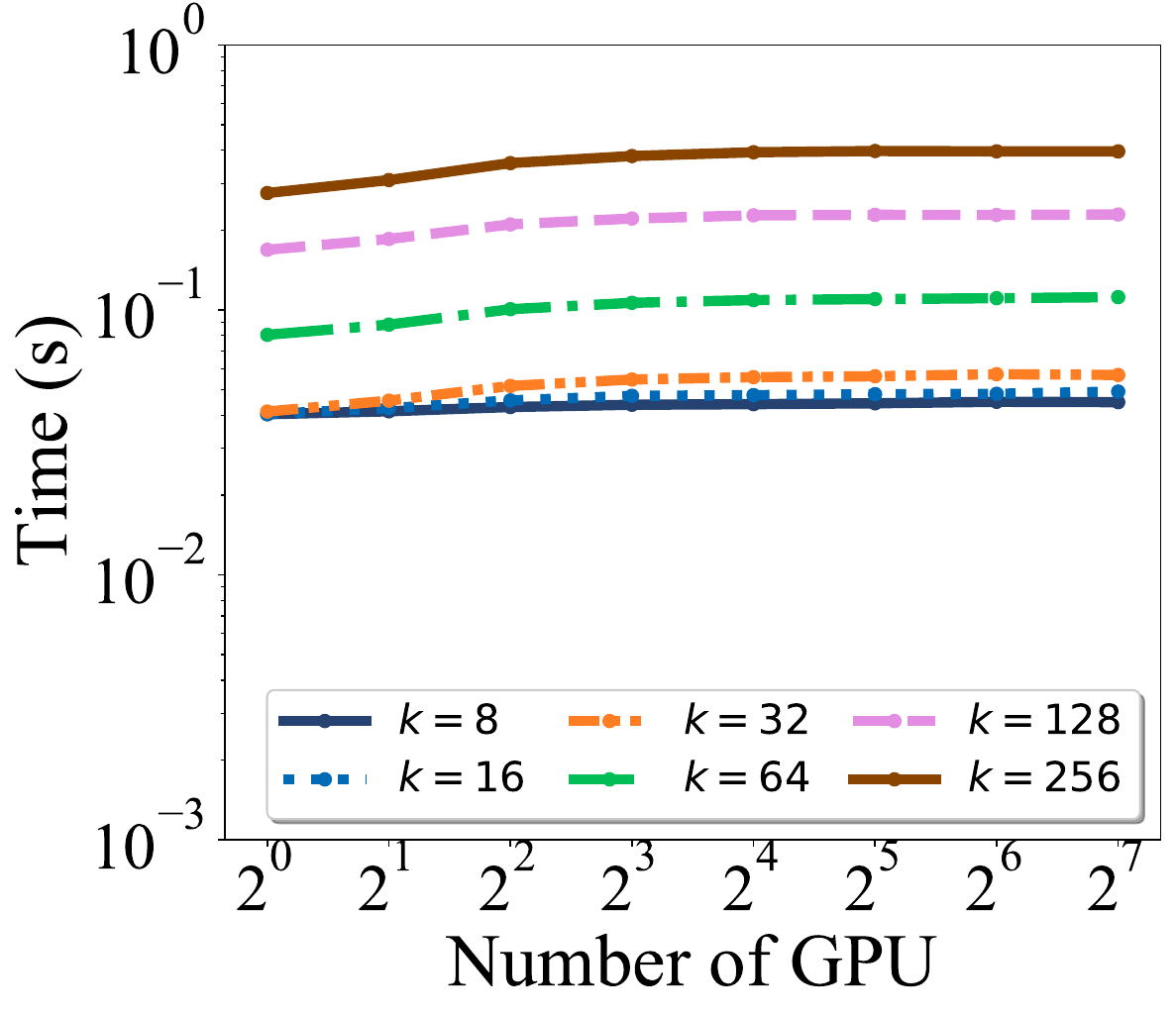}
        \subcaption{Dense \label{fig:weak_dense_n_t_k}}
    \end{minipage}
    \hspace{1mm}
    \begin{minipage}[b]{0.48\linewidth}
        \includegraphics[width=1\linewidth]{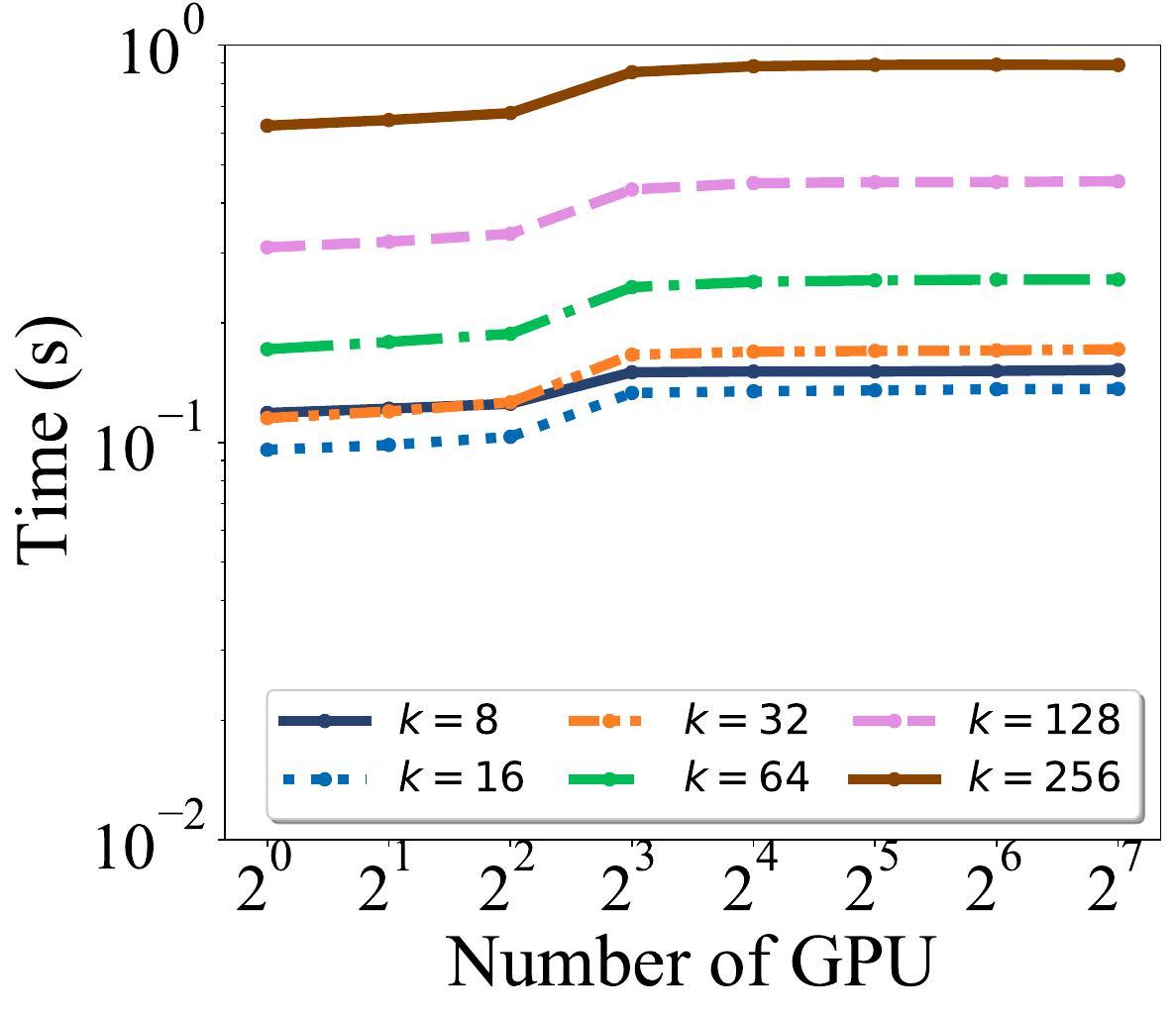}
        \subcaption{Sparse \label{fig:weak_sparse_n_t_k}}
    \end{minipage}
    \hspace{1mm}
    \begin{minipage}[b]{0.48\linewidth}
        \includegraphics[width=1\linewidth]{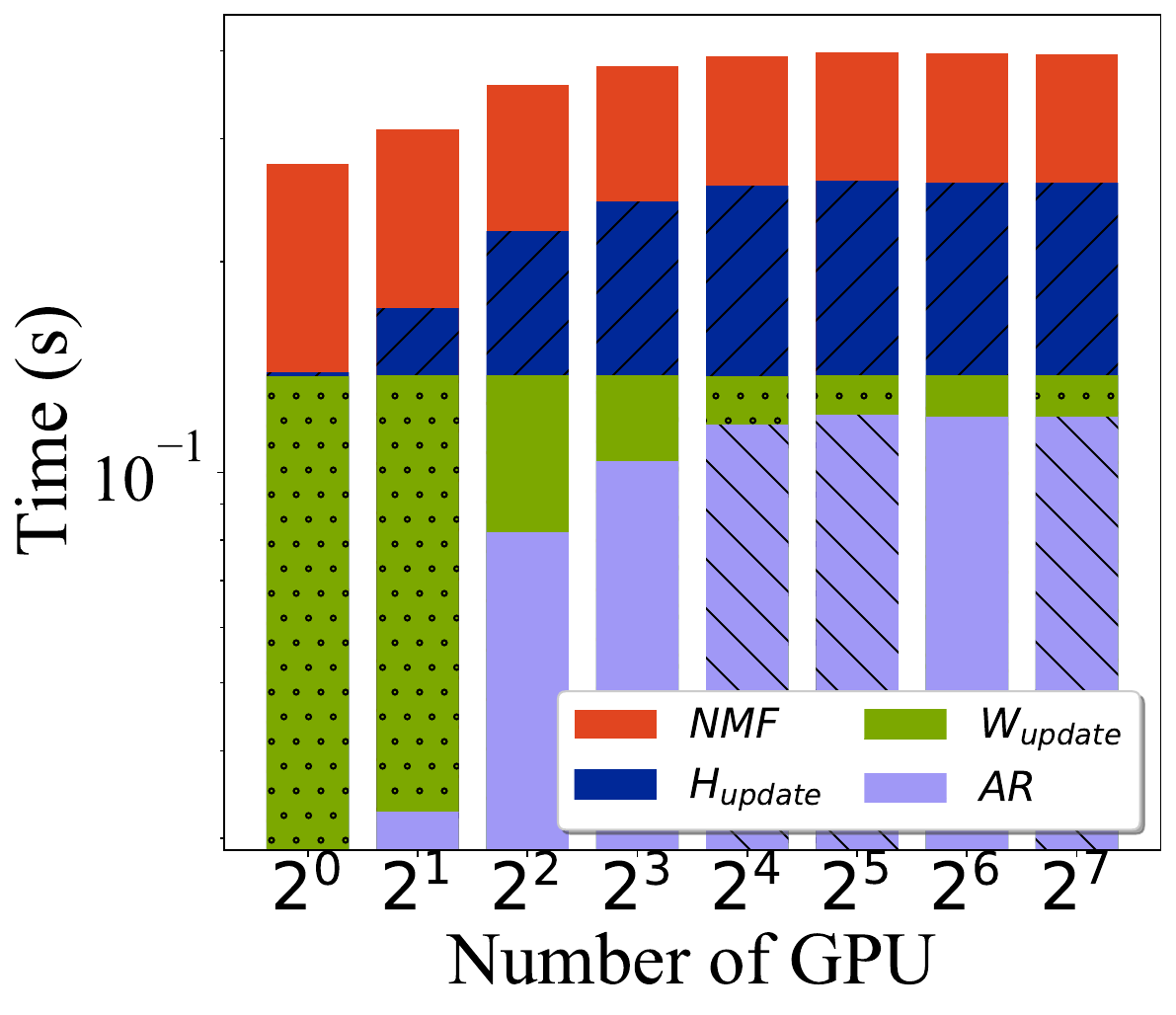}
        \subcaption{Dense \label{fig:weak_dense_n_t}}
    \end{minipage}
    \hspace{1mm}
    \begin{minipage}[b]{0.48\linewidth}
        \includegraphics[width=1\linewidth]{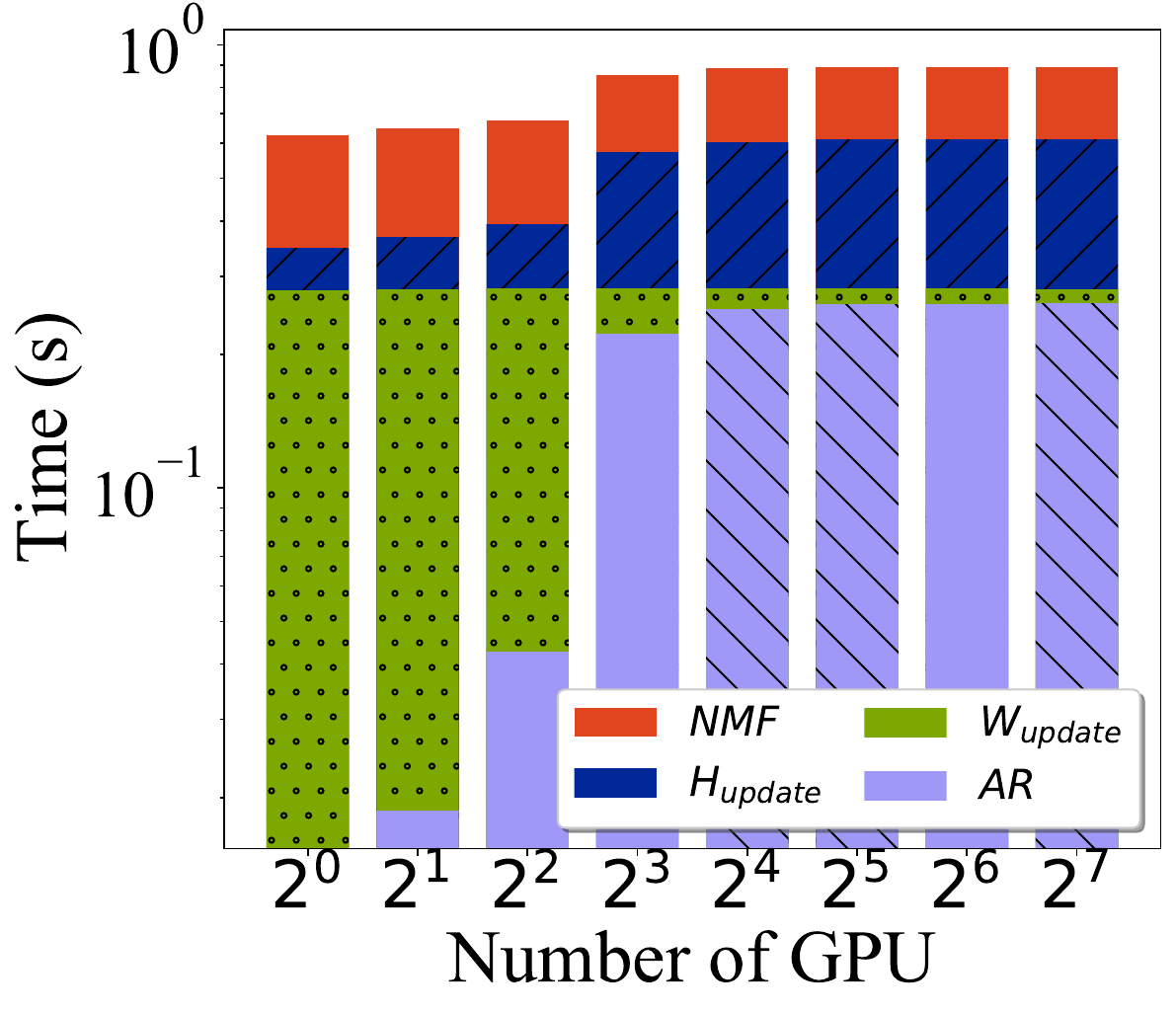}
        \subcaption{Sparse \label{fig:weak_sparse_n_t}}
    \end{minipage}
    \caption{ Results of weak scaling study performed on \emph{Kodiak}. NMF time vs number of GPU for various $k$ are respectively shown in (\ref{fig:weak_dense_n_t_k}) and (\ref{fig:weak_sparse_n_t_k}), for dense and sparse $\mat{A}$. For the case $k=8$, execution time of $H_{update}$, $W_{update}$ and All-reduce communication are compared in (\ref{fig:weak_dense_n_t}) and (\ref{fig:weak_sparse_n_t}), respectively for dense and sparse $\mat{A}$. }
    \label{fig:weak_scaling}
\end{figure}
Weak scaling results for cases with $k=8,16,32,64,128,256$ are shown Figure~\ref{fig:weak_dense_n_t_k}. Good weak scaling is indicated by constant NMF time with the increasing number of compute units, and this is observed only when $N>8$. The lack of scaling when $N <8$ can be explained using the breakdown of $H_{update}$, $W_{update}$ and combined $AR$ execution time for the case where $k=256$, shown in Figure~\ref{fig:weak_dense_n_t}. While $W_{update}$ maintains a perfect weak scaling at all $N$, $H_{update}$ is influenced by $AR$ communications time, which increases with GPU count. Communication grows with noticeable transitions indicating the use of slower channels. The first transition is from $N=1$ to $N=2$, indicating the beginning of $intra-node$ communication between GPUs on the same node. While growing with $N$, $intra-node$ communication remains a small portion of $W_{update}$ ( $\sim 10\%$). The next major transition occurs between $N=4$ and $N=8$, indicating the beginning of $inter-node$ communication, which quickly saturates to $\sim 40\%$ of $W_{update}$ by $N=32$.
Identical weak scaling is observed for cases where $A$ is sparse, as shown by plots in Figure~\ref{fig:weak_sparse_n_t_k}, and the explanation for lack of scaling when $N<8$ is consistent with the explanation given above for the case where $A$ is dense, as one can interpret from Figure~\ref{fig:weak_sparse_n_t}. 

\begin{figure}[t!]
\centering
    \begin{minipage}[b]{0.48\linewidth}
        \includegraphics[width=1\linewidth]{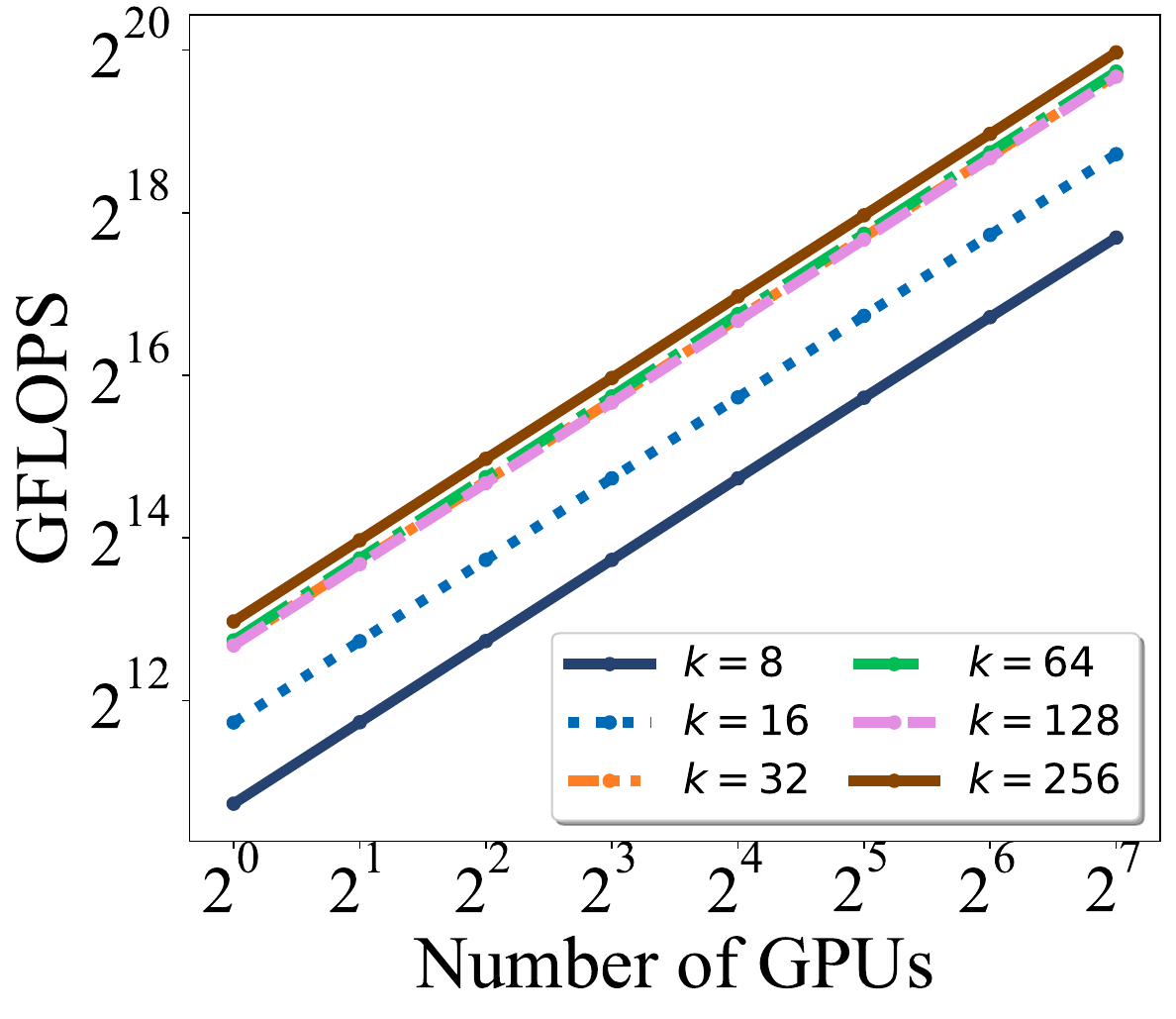}
        \subcaption{GFLOPS \label{fig:weak_FLOPS}}
    \end{minipage}
    \hspace{1mm}
    \begin{minipage}[b]{0.48\linewidth}
        \includegraphics[width=1\linewidth]{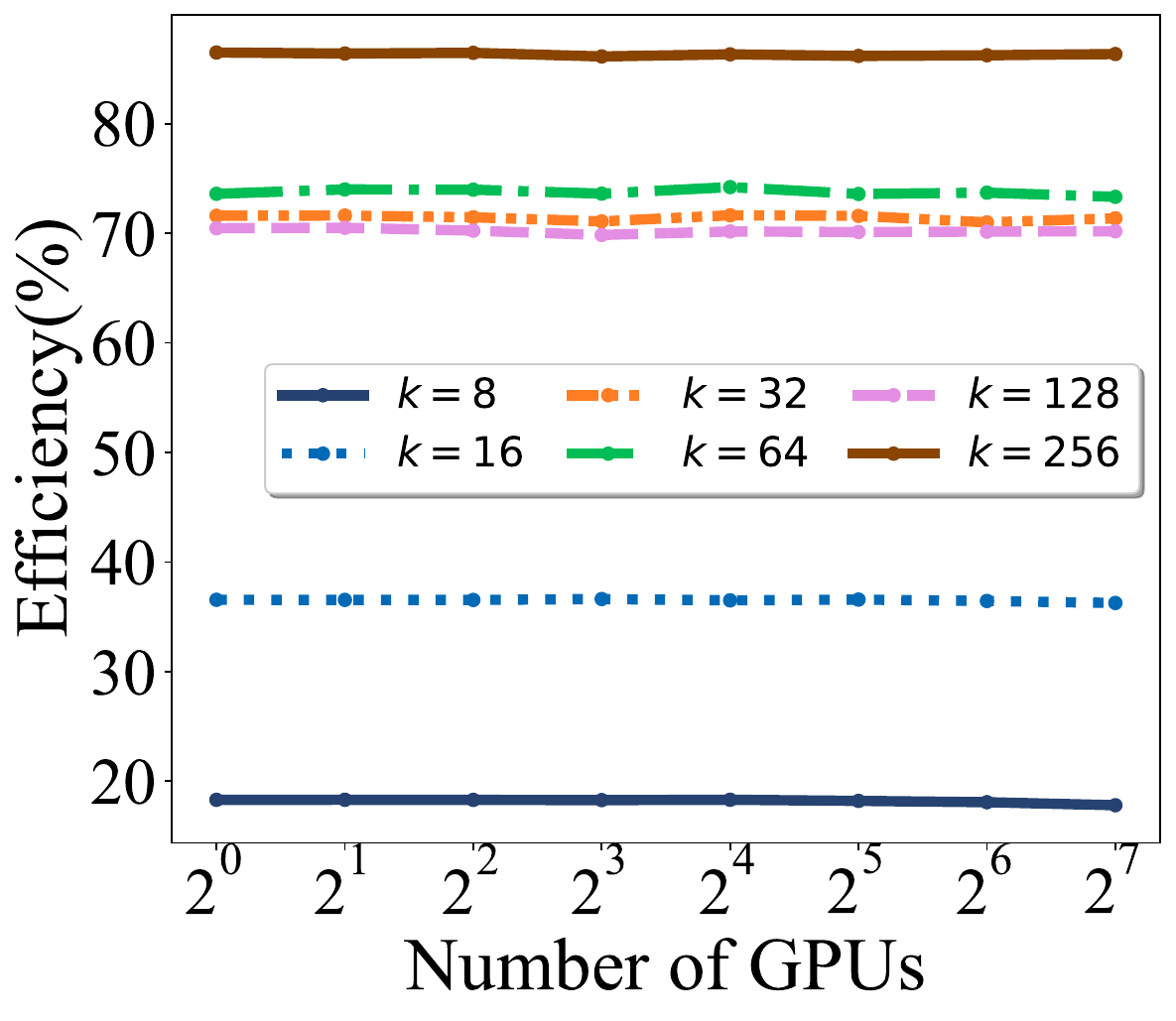}
        \subcaption{Efficiency \label{fig:weak_efficiency}}
    \end{minipage}

    \caption{FLOPS and Efficiency graph for weak scaling results for Kodiak Cluster  are shown  respectively in (a) and (b).}
    \label{fig:weak_scaling_eff}
    \vspace{-1em}
\end{figure}

In Figure~\ref{fig:weak_scaling_eff}, we display the GFLOPS and Efficiency results generated from our weak scaling experiments conducted on the Kodiak cluster.  Notably, GFLOPS shows a linear progression as GPU counts rise in Figure~\ref{fig:weak_FLOPS}, indicating an efficient distribution of computational workload across GPUs. Simultaneously, the consistent relationship of Efficiency with increasing GPU counts shown in Figure~\ref{fig:weak_efficiency} underscores the effective GPU utilization, thereby confirming our implementation's efficacy in maintaining performance at scale, specially for larger ranks(\emph{k}).

While all scaling results were obtained with \emph{RNMF}, similar results will be obtained with $\mat{A}^T$ using \emph{CNMF}.

\subsection{Scaling benchmark results on \emph{Big Data}}

It's important to note that as technology continues to evolve, the scale of data storage and processing capabilities will likely increase, leading to even more significant data sets in the future. 
“The world’s most valuable resource is no longer oil, but data1”~\cite{add1}. In national security and related research efforts, vast amounts of high-dimensional data are continuously being generated by massive computer simulations, large-scale experiments, surveillance systems, etc ~\cite{add2,add3}. For example, Stanford Synchrotron Radiation Lightsource experiments at SLAC laboratory for revealing the inner structure of materials at nanometer scales~\cite{add5,add6} and the Large Hadron Collider~\cite{add7} produce terabytes of data in minutes. Another example is the petabytes of data generated by mission-critical simulations~\cite{add8,add9,add10,add11,add12}. Exploration and analysis of such extra-large data mandates the development of novel machine learning (ML) approaches that are able to extract meaningful basic processes and fundamental features underlying the data~\cite{add13}.

Given our interest in exascale data, the proposed implementation was tested on a dense matrix of shape $[2618523648;32768]$ with a size of $\sim 340 TB $, and a sparse matrix of shape $[2.89 * 10^{12}, 1.05 * 10^6]$ with sparsity $10^{-6}$ and size of $\sim 11 EB$ ($\sim 34 TB$ when compressed in a sparse format). Benchmarks were performed on \emph{Summit}, with an allocation of 4096 nodes with 6 GPUs of 16 GB VRAM each, totaling a combined $~394 TB$ VRAM. While that is not enough to efficiently factorize either of the two matrices, we chose to cache $A$ and co-factors and batch the compute of heavy, intermediate products (OOM-0). This way, we can reduce performance loss by avoiding unnecessary data transfers from host to device and vice-versa. 

\begin{figure}[t!]
\centering
    \begin{minipage}[b]{0.48\linewidth}
        \includegraphics[width=1\linewidth]{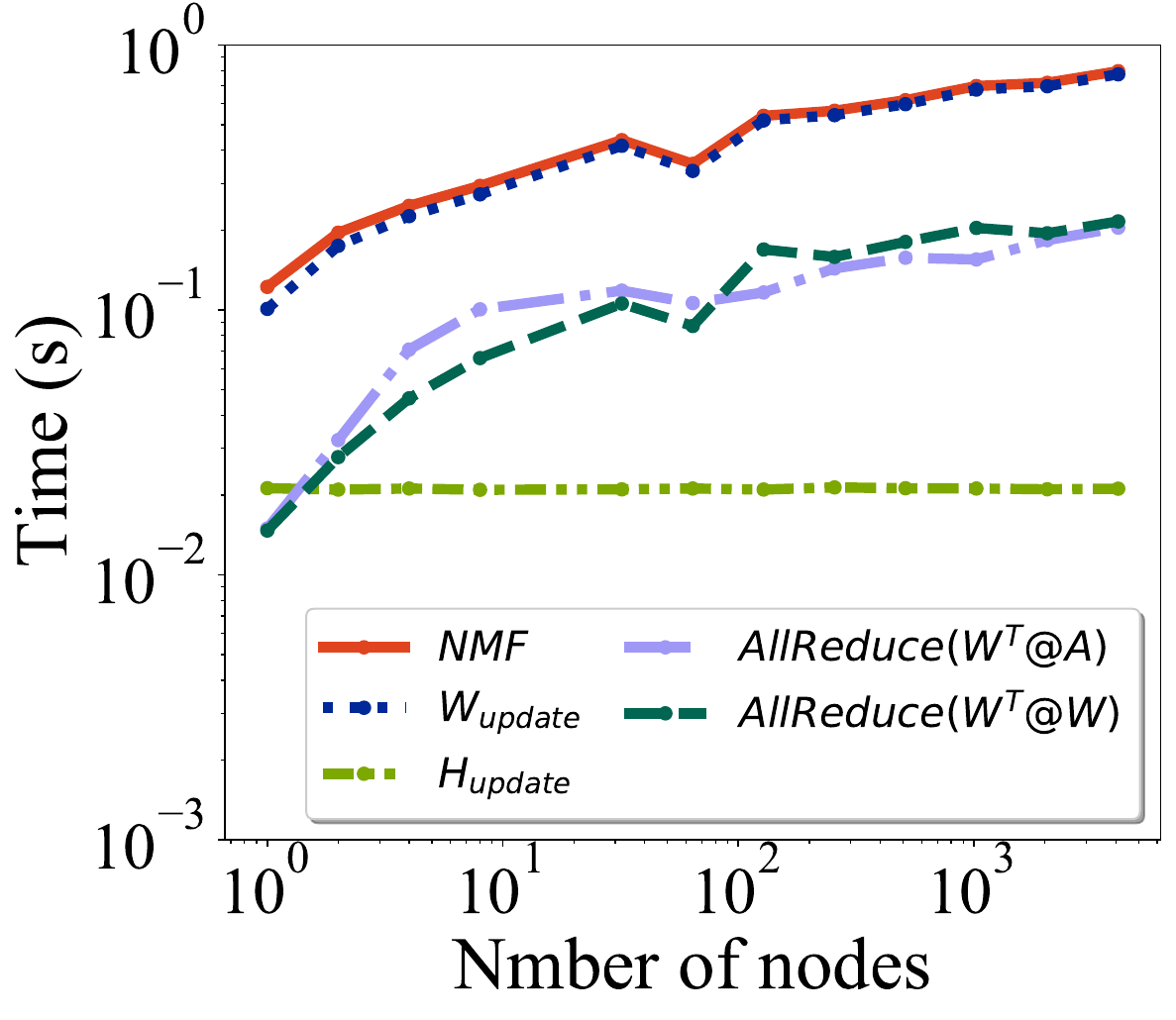}
        \subcaption{Dense \label{fig:weak_dense_XL}}
    \end{minipage}
    \hspace{1mm}
    \begin{minipage}[b]{0.48\linewidth}
        \includegraphics[width=1\linewidth]{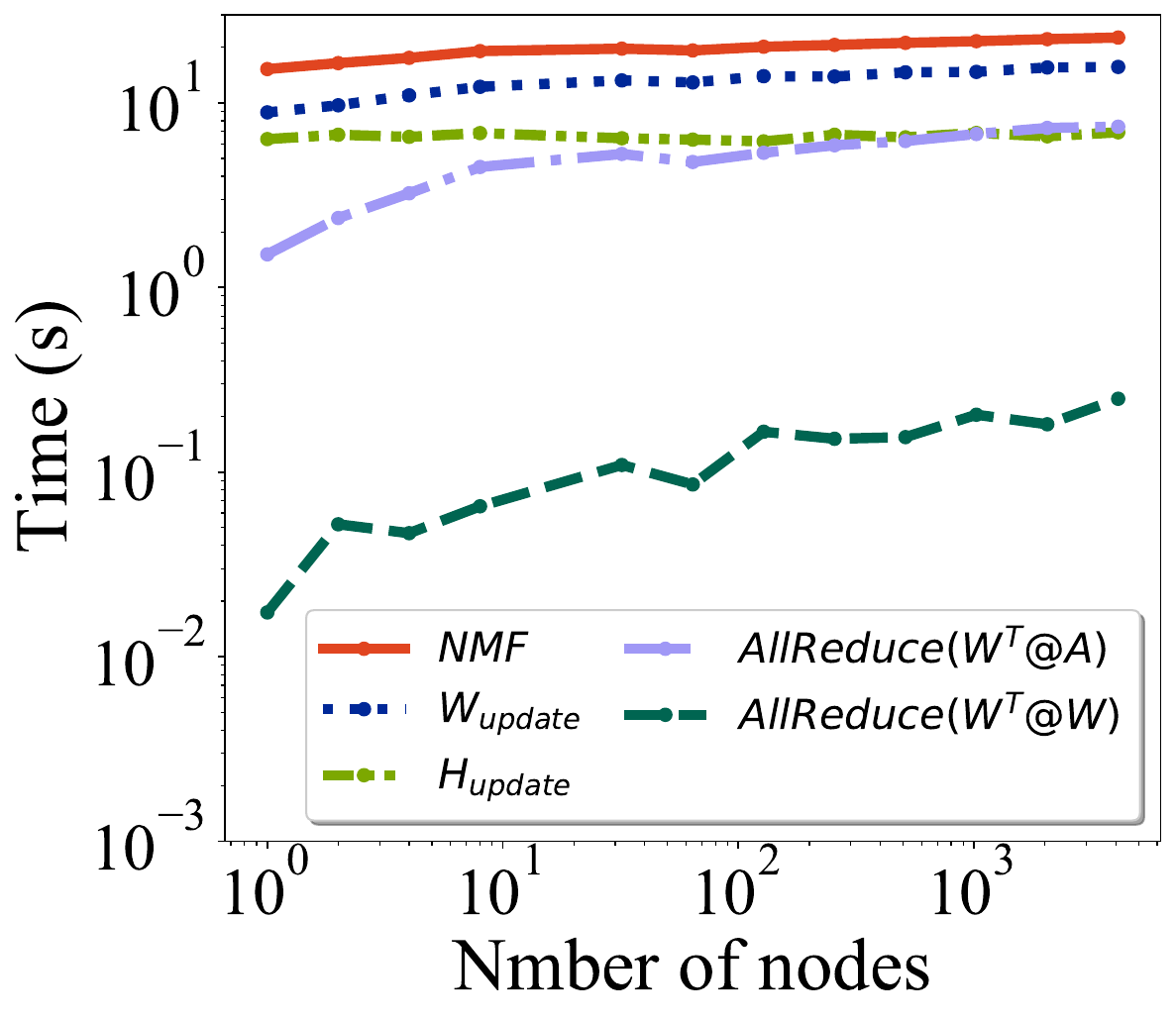}
        \subcaption{Sparse \label{fig:weak_sparse_XL}}
    \end{minipage}

    \caption{Results of weak scaling study for dense and sparse $\mat{A}$ performed on \emph{Summit}  are shown  respectively in (a) and (b).}
    \label{fig:weak_scaling_XL}
    \vspace{-1em}
\end{figure}

On the one hand, the weak scaling benchmark results for the dense array are reported in Figure~\ref{fig:weak_dense_XL}. The $H_{update}$ is shown with a perfect weak scaling, while the $W_{update}$ is shown not to scale appropriately. Loss of scaling in the $W_{update}$ is a consequence of the high communication cost associated with the \emph{All-reduce} of $W^TA$ and $W^TW$, which combined, make up a substantial portion of the $W_{update}$. The total NMF time, in turn, is significantly affected by the $W_{update}$, which takes about one order of magnitude more time to execute than the $H_{update}$.
On the other, the weak scaling benchmark results for the sparse array, reported in Figure~\ref{fig:weak_sparse_XL}, indicate both $W_{update}$ and $H_{update}$ to have an excellent weak scaling. The $AR(W^TW)$ is similar in both cases, as $W^TW$ is of shape $k \times k$, but the $AR(W^TA)$ is two orders of magnitude higher in the case of the spare dataset, proportional to $n$ which is also two orders of magnitude higher. Unlike in the case of the dense array, the communication cost associated with the $AR(W^TA)$ and $AR(W^TW)$, although higher, are not significant enough to affect the $W_{update}$, consequently do not affect the overall scaling of the NMF.

\subsection{Benchmark results on out-of-memory problems}

\begin{figure}[t!]
\centering
    \begin{minipage}[b]{0.48\linewidth}
        \includegraphics[width=1\linewidth]{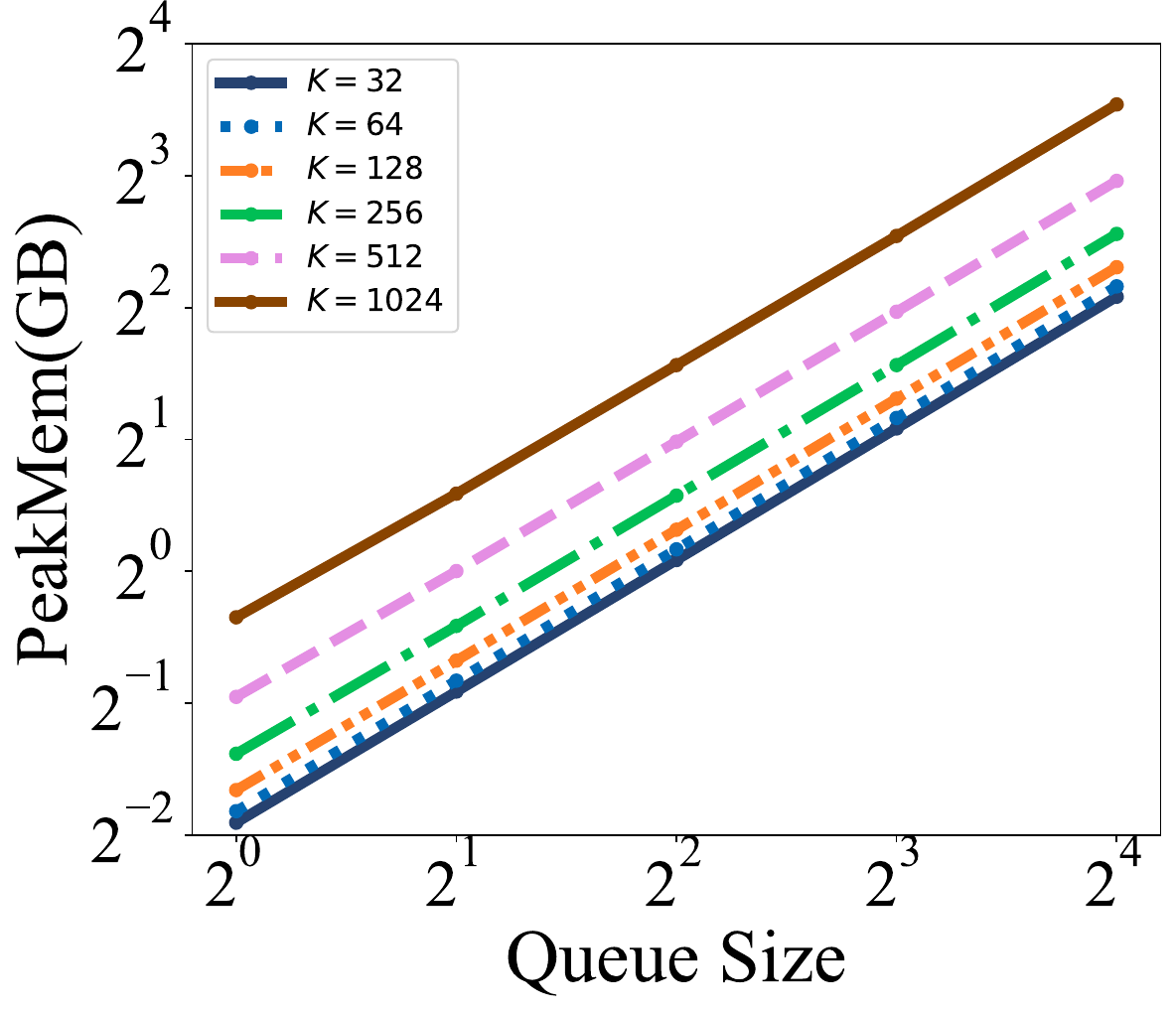}
        \subcaption{OOM Memory peak \label{fig:oom_mem_nmf}}
    \end{minipage}
    \hspace{1mm}
    \begin{minipage}[b]{0.48\linewidth}
        \includegraphics[width=1\linewidth]{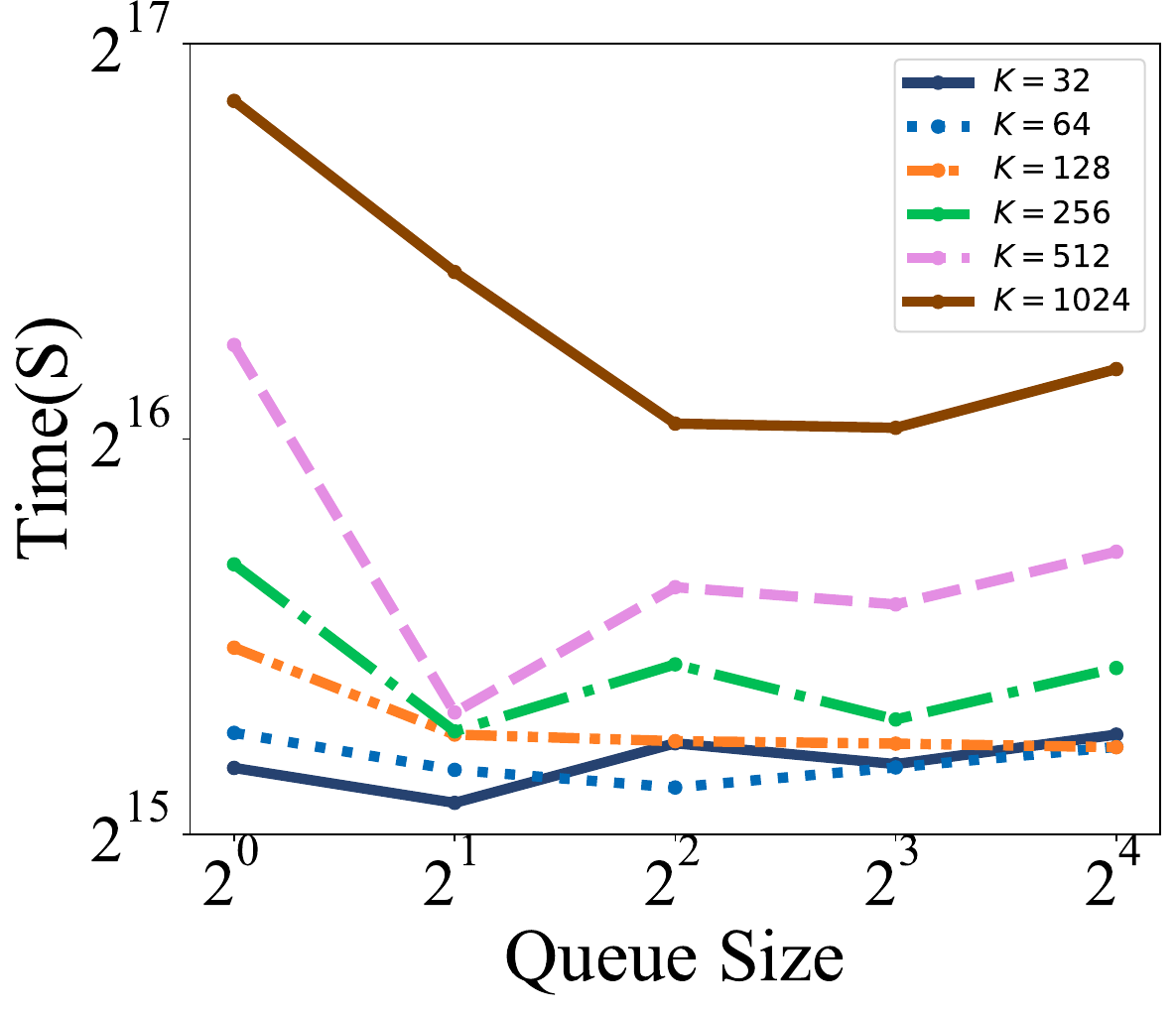}
        \subcaption{ OOM  execution time\label{fig:oom_time_nmf}}
    \end{minipage}

    \caption{Results of Out of memory  NMF benchmarks on \emph{Chicoma} showing  (a) NMF peak memory vs queue sizes for different $k$, and (b) NMF execution time vs queue sizes for different $k$.}
    \label{fig:OOM_benchmark}
    \vspace{-1em}
\end{figure}

Next, we assess the effectiveness of the proposed batching technique for OOM scenarios and the use of the CUDA stream queues to reduce communication in Algorithm~ ~\ref{alg:batched_RNMF}. To this end, the proposed implementation is tested in an OOM-1 scenario, where a matrix of shape [524288, 4096] is factorized  for $k=[32,64,128,256,512,1024]$. Smaller array $\mat{H}$ is cached on GPU memory, and large arrays $\mat{A}$ and $\mat{W}$  are stored on the host and batched to GPU as needed. For this experiment, the number of iterations in Algorithm~ ~\ref{alg:batched_RNMF}(line 4) fixed to $max\_iters=100$, and the number of batches is fixed to $n_b=32$. Given the size of $A$ in single precision is $S_A=8$GB, the resulting batch size is $S_B=p\times n \sim 0.25$GB. The GPU peak memory utilization and NMF execution time for the 100 iteration, vs queue size, are respectively reported in Figure~\ref{fig:oom_mem_nmf} and Figure~\ref{fig:oom_time_nmf}.

In Figure~\ref{fig:oom_mem_nmf}, the peak memory utilization measured when $q_s=1$ is $S_{nmf} \sim 0.267GB$ which is close to the estimated memory complexity of $\mathcal{O}(p \times n \times q_s) \approx 0.25$GB in  section~\ref{subsec:dist_batching}, and which is a very big saving, $\sim 1/100X$, compared to the estimated $S_{NMF}~3 \times S_A \approx 24$ GB require by a normal implementation. This memory complexity is maintained for all $k$ values and all queue sizes as indicated by the lines with the same slope $\sim 0.267$ in Figure~\ref{fig:oom_mem_nmf}. The increase in peak memory with increasing $k$ for any given queue size is explained by the increase in the size of the arrays cached on GPU ($\mat{H}$), as well as the increase in the size of the computed intermediate products (see Figure~\ref{fig:MU_batch_row_algo2}. Similarly, for each $k$ value, we note an increase in peak memory utilization with the increasing number of batches which is simply explained by the aggregated memory utilization from the concurrent streams. While from this figure, it seems unproductive to use larger stream queue sizes due to the increase in peak memory utilization, the benefits of such design choice are explained in the execution benchmark results reported in Figure~\ref{fig:oom_time_nmf}.

From Figure~\ref{fig:oom_time_nmf}, we first see that it is, in all cases, a good idea to choose a queue size $q_s>1$ if one wants to speed up the \emph{NMFk} execution time. This is explained by using large stream queue sizes makes more streams available to overlap memory copies, all-reduce communications, and compute concurrently. It is, however, not the case that more streams will always make this process better, as we can see it not being the case when $q_s=16$, where the \emph{NMFk} execution time is not optimum for any $k$ value. This is explained by the fact that CUDA core counts are limited and that some streams will block and wait when all cores are busy processing other streams, causing load-balancing delays. Consequently, it is crucial to fine-tune $q_s$ for a given batch size and $k$ to obtain optimal performance.

\subsection{Validation of the model selection capability}

\begin{figure}[ht!] 
    \begin{subfigure}[b]{0.49\textwidth}
       \centering
        \includegraphics[width=1\textwidth]{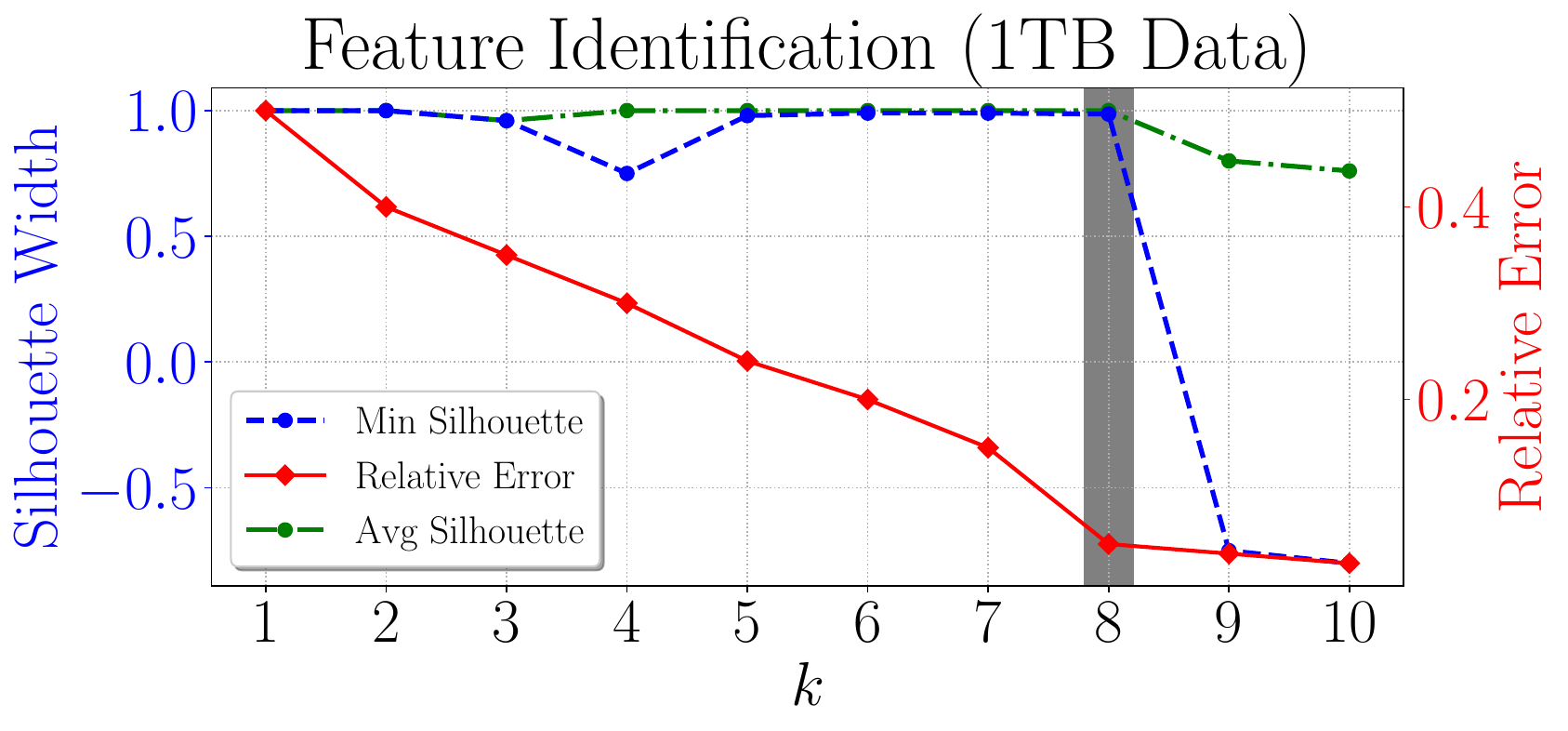}
        \caption{} \label{fig:sill_TB}
    \end{subfigure}
    \begin{subfigure}[b]{0.49\textwidth}
        \centering
        \includegraphics[width=.7\textwidth]{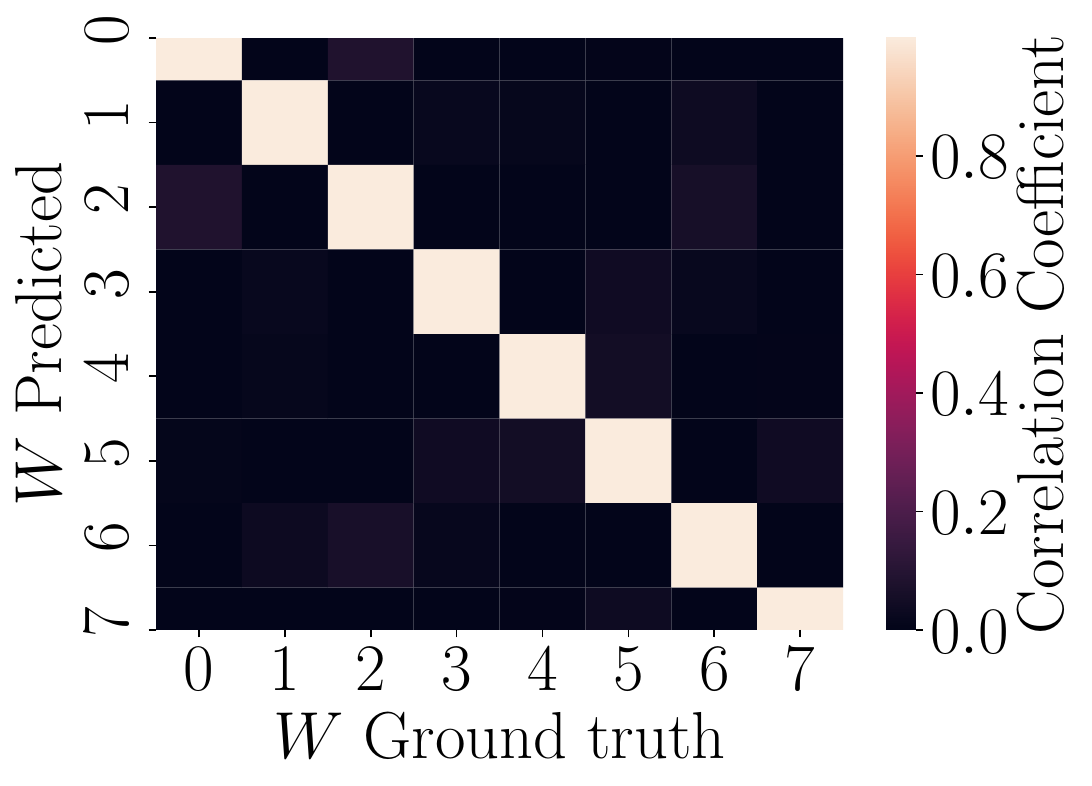}        \caption{} \label{fig:cf_mtx}
    \end{subfigure}
    \caption{(a) Estimation of number of hidden features(k=8) through Silhouette analysis \cite{chennupati2020distributed}. (b) Pearson correlation between columns of ground truth W and reconstructed W.}
    
\end{figure}
\label{sec:Experiments}
To demonstrate the correctness of the proposed algorithm on big synthetic datasets, we first integrate our \emph{pyDNMF-GPU} with the existing model selection algorithm 
  \emph{pyDNMFk}~\cite{bhattarai2021pydnmfk}. Then, we determine the number of latent features on a synthetic terabyte size matrix (with a predetermined number of features) and show that estimation is performed correctly. We generate a random matrix of dimensions $8388608\times 32768$ as a product of two random matrices, $\mat{W}$ and $\mat{H}$, with a latent feature count of $k=8$. We construct $\mat{W}$ with Gaussian features with different statistical means. The \textit{pyDNMFk-GPU} silhouette analysis corresponding to this decomposition is shown in Figure~\ref{fig:sill_TB} and the correctness of features is shown with confusion matrix in Figure~\ref{fig:cf_mtx}. pyDNMFk-GPU estimates $k=8$ as the minimum Silhouette score is high and relative error is low. For $k>8$, the minimum silhouette score drops suddenly as the solutions begin to fit the noise Figure~\ref{fig:sill_TB}. Figure~\ref{fig:cf_mtx} shows a Pearson correlation matrix that illustrates a large correlation between the features of ground truth \emph{$\mat{W}$ Ground truth} and the corresponding pyDNMFk-GPU extracted $\mat{W} Predicted$ for $k=8$. The analysis took approximately 1 hour to correctly estimate the latent features on \emph{Kodiak}. The average reconstruction error for the data is $\sim 4\%$ with the Frobenius norm objective and MU update optimization. Our experiment demonstrates that \textit{pyDNMFk-GPU} correctly estimates the number of latent features in addition to its scalability for large datasets as demonstrated in previous sections.

\section{Conclusion}
\label{sec:Conclusion}

In summary, we demonstrated a novel scalable and portable framework, \textit{pyDNMFk-GPU}, for non-negative matrix factorization based on custom multiplicative updates, with automatic determination of the number of latent features on Exa-scale data. Scalability of the framework was demonstrated via strong and weak scaling benchmarks, and speedup gains on GPU over CPU were found to vary with $k$ and to increase with the size of the HPC system. The efficacy of the proposed tiling technique was demonstrated through the OOM-0 problem by factorizing a dense dataset of 340TB and a sparse dataset of size 11EB, where the implementation was found to have good week scaling on upto to 25k GPU. We also demonstrated the efficacity of the proposed batching technique along with the importance of using CUDA streams by solving \emph{OOM-1} problem, where memory complexity was shown to be of the $\mathcal{O}(p \times n \times q_s)$, resulting in a significant saving of $\sim 100X$ smaller peak memory utilization in some cases. The automatic model selection capability was verified by correctly decomposing large synthetic data with a predetermined number of latent features and factors.

\section*{Declarations}

\subsection*{Competing interests}
The authors declare that they have no competing interest.



\subsection*{Acknowledgement}
This research used resources  of Los Alamos National Laboratory
Institutional Computing Program, supported by the U.S.
Department of Energy National Nuclear Security Administration under Contract No. 89233218CNA000001 and 
the Oak Ridge Leadership Computing Facility at the Oak Ridge National Laboratory under Director’s Discretionary allocation \#CSC456, which is supported by the Office of Science of the U.S. Department of Energy under Contract No. DE-AC05-00OR22725.

 \subsection*{Funding}
This research was funded by DOE National Nuclear Security Administration  (NNSA)  -  Office  of  Defense  Nuclear  Non-proliferation  R\&D  (NA-22) and by U.S. Department of Energy National Nuclear Security Administration under Contract No. DE-AC52-06NA25396 and through LANL laboratory directed research and development (LDRD) grant 20190020DR.

\subsection*{Availability of data and materials} 
The code and the benchmark results used in this paper will be available at https://github.com/lanl/pyDNMFk .

%
%
%

\bibliography{main}


\begin{thebibliography}{51}
\ifx \bisbn   \undefined \def \bisbn  #1{ISBN #1}\fi
\ifx \binits  \undefined \def \binits#1{#1}\fi
\ifx \bauthor  \undefined \def \bauthor#1{#1}\fi
\ifx \batitle  \undefined \def \batitle#1{#1}\fi
\ifx \bjtitle  \undefined \def \bjtitle#1{#1}\fi
\ifx \bvolume  \undefined \def \bvolume#1{\textbf{#1}}\fi
\ifx \byear  \undefined \def \byear#1{#1}\fi
\ifx \bissue  \undefined \def \bissue#1{#1}\fi
\ifx \bfpage  \undefined \def \bfpage#1{#1}\fi
\ifx \blpage  \undefined \def \blpage #1{#1}\fi
\ifx \burl  \undefined \def \burl#1{\textsf{#1}}\fi
\ifx \doiurl  \undefined \def \doiurl#1{\url{https://doi.org/#1}}\fi
\ifx \betal  \undefined \def \betal{\textit{et al.}}\fi
\ifx \binstitute  \undefined \def \binstitute#1{#1}\fi
\ifx \binstitutionaled  \undefined \def \binstitutionaled#1{#1}\fi
\ifx \bctitle  \undefined \def \bctitle#1{#1}\fi
\ifx \beditor  \undefined \def \beditor#1{#1}\fi
\ifx \bpublisher  \undefined \def \bpublisher#1{#1}\fi
\ifx \bbtitle  \undefined \def \bbtitle#1{#1}\fi
\ifx \bedition  \undefined \def \bedition#1{#1}\fi
\ifx \bseriesno  \undefined \def \bseriesno#1{#1}\fi
\ifx \blocation  \undefined \def \blocation#1{#1}\fi
\ifx \bsertitle  \undefined \def \bsertitle#1{#1}\fi
\ifx \bsnm \undefined \def \bsnm#1{#1}\fi
\ifx \bsuffix \undefined \def \bsuffix#1{#1}\fi
\ifx \bparticle \undefined \def \bparticle#1{#1}\fi
\ifx \barticle \undefined \def \barticle#1{#1}\fi
\bibcommenthead
\ifx \bconfdate \undefined \def \bconfdate #1{#1}\fi
\ifx \botherref \undefined \def \botherref #1{#1}\fi
\ifx \url \undefined \def \url#1{\textsf{#1}}\fi
\ifx \bchapter \undefined \def \bchapter#1{#1}\fi
\ifx \bbook \undefined \def \bbook#1{#1}\fi
\ifx \bcomment \undefined \def \bcomment#1{#1}\fi
\ifx \oauthor \undefined \def \oauthor#1{#1}\fi
\ifx \citeauthoryear \undefined \def \citeauthoryear#1{#1}\fi
\ifx \endbibitem  \undefined \def \endbibitem {}\fi
\ifx \bconflocation  \undefined \def \bconflocation#1{#1}\fi
\ifx \arxivurl  \undefined \def \arxivurl#1{\textsf{#1}}\fi
\csname PreBibitemsHook\endcsname

\bibitem{lee1999learning}
\begin{barticle}
\bauthor{\bsnm{Lee}, \binits{D.D.}},
\bauthor{\bsnm{Seung}, \binits{H.S.}}:
\batitle{{Learning the parts of objects by non-negative matrix factorization}}.
\bjtitle{Nature}
\bvolume{401}(\bissue{6755}),
\bfpage{788}--\blpage{791}
(\byear{1999})
\end{barticle}
\endbibitem

\bibitem{cichocki2009nonnegative}
\begin{botherref}
\oauthor{\bsnm{Cichocki}, \binits{A.}},
\oauthor{\bsnm{Zdunek}, \binits{R.}},
\oauthor{\bsnm{Phan}, \binits{A.H.}},
\oauthor{\bsnm{Amari}, \binits{S.-i.}}:
{Nonnegative matrix and tensor factorizations: applications to exploratory
  multi-way data analysis and blind source separation}
(2009)
\end{botherref}
\endbibitem

\bibitem{everett2013introduction}
\begin{botherref}
\oauthor{\bsnm{Everett}, \binits{B.}}:
{An introduction to latent variable models}
(2013)
\end{botherref}
\endbibitem

\bibitem{alexandrov2013deciphering}
\begin{barticle}
\bauthor{\bsnm{Alexandrov}, \binits{L.B.}},
\bauthor{\bsnm{Nik-Zainal}, \binits{S.}},
\bauthor{\bsnm{Wedge}, \binits{D.C.}},
\bauthor{\bsnm{Campbell}, \binits{P.J.}},
\bauthor{\bsnm{Stratton}, \binits{M.R.}}:
\batitle{{Deciphering signatures of mutational processes operative in human
  cancer}}.
\bjtitle{Cell reports}
\bvolume{3}(\bissue{1}),
\bfpage{246}--\blpage{259}
(\byear{2013})
\end{barticle}
\endbibitem

\bibitem{alexandrov2020source}
\begin{botherref}
\oauthor{\bsnm{Alexandrov}, \binits{B.S.}},
\oauthor{\bsnm{Alexandrov}, \binits{L.B.}},
\oauthor{\bsnm{Iliev}, \binits{F.}},
\oauthor{\bsnm{Stanev}, \binits{V.G.}},
\oauthor{\bsnm{Vesselinov}, \binits{V.}}:
Source identification by non-negative matrix factorization combined with
  semi-supervised clustering.
Google Patents.
US Patent 10,776,718
(2020)
\end{botherref}
\endbibitem

\bibitem{chennupati2020distributed}
\begin{botherref}
\oauthor{\bsnm{Chennupati}, \binits{G.}},
\oauthor{\bsnm{Vangara}, \binits{R.}},
\oauthor{\bsnm{Skau}, \binits{E.}},
\oauthor{\bsnm{Djidjev}, \binits{H.}},
\oauthor{\bsnm{Alexandrov}, \binits{B.}}:
{Distributed non-negative matrix factorization with determination of the number
  of latent features}.
The Journal of Supercomputing,
1--31
(2020)
\end{botherref}
\endbibitem

\bibitem{bhattarai2021pydnmfk}
\begin{botherref}
\oauthor{\bsnm{Bhattarai}, \binits{M.}},
\oauthor{\bsnm{Nebgen}, \binits{B.}},
\oauthor{\bsnm{Skau}, \binits{E.}},
\oauthor{\bsnm{Eren}, \binits{M.}},
\oauthor{\bsnm{Chennupati}, \binits{G.}},
\oauthor{\bsnm{Vangara}, \binits{R.}},
\oauthor{\bsnm{Djidjev}, \binits{H.}},
\oauthor{\bsnm{Patchett}, \binits{J.}},
\oauthor{\bsnm{Ahrens}, \binits{J.}},
\oauthor{\bsnm{ALexandrov}, \binits{B.}}:
pyDNMFk: Python Distributed Non Negative Matrix Factorization.
GitHub
(2021).
\doiurl{10.5281/zenodo.4722448}
\end{botherref}
\endbibitem

\bibitem{vangara2021finding}
\begin{botherref}
\oauthor{\bsnm{Vangara}, \binits{R.}},
\oauthor{\bsnm{Bhattarai}, \binits{M.}},
\oauthor{\bsnm{Skau}, \binits{E.}},
\oauthor{\bsnm{Chennupati}, \binits{G.}},
\oauthor{\bsnm{Djidjev}, \binits{H.}},
\oauthor{\bsnm{Tierney}, \binits{T.}}, et al.:
{Finding the Number of Latent Topics with Semantic Non-negative Matrix
  Factorization}.
IEEE Access,
117217--117231
(2021)
\end{botherref}
\endbibitem

\bibitem{alexandrov2013signatures}
\begin{barticle}
\bauthor{\bsnm{Alexandrov}, \binits{L.B.}},
\bauthor{\bsnm{Nik-Zainal}, \binits{S.}},
\bauthor{\bsnm{Wedge}, \binits{D.C.}},
\bauthor{\bsnm{Aparicio}, \binits{S.A.}},
\bauthor{\bsnm{Behjati}, \binits{S.}},
\bauthor{\bsnm{Biankin}, \binits{A.V.}},
\bauthor{\bsnm{Bignell}, \binits{G.R.}},
\bauthor{\bsnm{Bolli}, \binits{N.}},
\bauthor{\bsnm{Borg}, \binits{A.}},
\bauthor{\bsnm{B{\o}rresen-Dale}, \binits{A.-L.}}, \betal:
\batitle{{Signatures of mutational processes in human cancer}}.
\bjtitle{Nature}
\bvolume{500}(\bissue{7463}),
\bfpage{415}
(\byear{2013})
\end{barticle}
\endbibitem

\bibitem{alexandrov2020repertoire}
\begin{barticle}
\bauthor{\bsnm{Alexandrov}, \binits{L.B.}},
\bauthor{\bsnm{Kim}, \binits{J.}},
\bauthor{\bsnm{Haradhvala}, \binits{N.J.}},
\bauthor{\bsnm{Huang}, \binits{M.N.}},
\bauthor{\bsnm{Ng}, \binits{A.W.T.}},
\bauthor{\bsnm{Wu}, \binits{Y.}},
\bauthor{\bsnm{Boot}, \binits{A.}},
\bauthor{\bsnm{Covington}, \binits{K.R.}},
\bauthor{\bsnm{Gordenin}, \binits{D.A.}},
\bauthor{\bsnm{Bergstrom}, \binits{E.N.}}, \betal:
\batitle{{The repertoire of mutational signatures in human cancer}}.
\bjtitle{Nature}
\bvolume{578}(\bissue{7793}),
\bfpage{94}--\blpage{101}
(\byear{2020})
\end{barticle}
\endbibitem

\bibitem{vangara2020semantic}
\begin{botherref}
\oauthor{\bsnm{Vangara}, \binits{R.}},
\oauthor{\bsnm{Skau}, \binits{E.}},
\oauthor{\bsnm{Chennupati}, \binits{G.}},
\oauthor{\bsnm{Djidjev}, \binits{H.}},
\oauthor{\bsnm{Tierney}, \binits{T.}},
\oauthor{\bsnm{Smith}, \binits{J.P.}},
\oauthor{\bsnm{Bhattarai}, \binits{M.}},
\oauthor{\bsnm{Stanev}, \binits{V.G.}},
\oauthor{\bsnm{Alexandrov}, \binits{B.S.}}:
{Semantic nonnegative matrix factorization with automatic model determination
  for topic modeling},
328--335
(2020).
IEEE
\end{botherref}
\endbibitem

\bibitem{bhattarai2020distributed}
\begin{bchapter}
\bauthor{\bsnm{Bhattarai}, \binits{M.}},
\bauthor{\bsnm{Chennupati}, \binits{G.}},
\bauthor{\bsnm{Skau}, \binits{E.}},
\bauthor{\bsnm{Vangara}, \binits{R.}},
\bauthor{\bsnm{Djidjev}, \binits{H.}},
\bauthor{\bsnm{Alexandrov}, \binits{B.S.}}:
\bctitle{{Distributed non-negative tensor train decomposition}}.
In: \bbtitle{2020 IEEE High Performance Extreme Computing Conference (HPEC)},
pp. \bfpage{1}--\blpage{10}
(\byear{2020}).
\bcomment{IEEE}
\end{bchapter}
\endbibitem

\bibitem{alexandrov2019nonnegative}
\begin{barticle}
\bauthor{\bsnm{Alexandrov}, \binits{B.S.}},
\bauthor{\bsnm{Stanev}, \binits{V.G.}},
\bauthor{\bsnm{Vesselinov}, \binits{V.V.}},
\bauthor{\bsnm{Rasmussen}, \binits{K.{\O}.}}:
\batitle{{Nonnegative tensor decomposition with custom clustering for
  microphase separation of block copolymers}}.
\bjtitle{Statistical Analysis and Data Mining: The ASA Data Science Journal}
\bvolume{12}(\bissue{4}),
\bfpage{302}--\blpage{310}
(\byear{2019})
\end{barticle}
\endbibitem

\bibitem{s.20211055}
\begin{bchapter}
\bauthor{\bsnm{Pulido}, \binits{J.}},
\bauthor{\bsnm{Patchett}, \binits{J.}},
\bauthor{\bsnm{Bhattarai}, \binits{M.}},
\bauthor{\bsnm{Alexandrov}, \binits{B.}},
\bauthor{\bsnm{Ahrens}, \binits{J.}}:
\bctitle{{Selection of Optimal Salient Time Steps by Non-negative Tucker Tensor
  Decomposition}}.
In: \beditor{\bsnm{Agus}, \binits{M.}},
\beditor{\bsnm{Garth}, \binits{C.}},
\beditor{\bsnm{Kerren}, \binits{A.}} (eds.)
\bbtitle{EuroVis 2021 - Short Papers}.
\bpublisher{The Eurographics Association}, \blocation{???}
(\byear{2021}).
\doiurl{10.2312/evs.20211055}
\end{bchapter}
\endbibitem

\bibitem{bhattarai2022distributed}
\begin{botherref}
\oauthor{\bsnm{Bhattarai}, \binits{M.}},
\oauthor{\bsnm{Kharat}, \binits{N.}},
\oauthor{\bsnm{Skau}, \binits{E.}},
\oauthor{\bsnm{Nebgen}, \binits{B.}},
\oauthor{\bsnm{Djidjev}, \binits{H.}},
\oauthor{\bsnm{Rajopadhye}, \binits{S.}},
\oauthor{\bsnm{Smith}, \binits{J.P.}},
\oauthor{\bsnm{Alexandrov}, \binits{B.}}:
Distributed non-negative rescal with automatic model selection for exascale
  data.
arXiv preprint arXiv:2202.09512
(2022)
\end{botherref}
\endbibitem

\bibitem{pyDRESCALk}
\begin{botherref}
\oauthor{\bsnm{Bhattarai}, \binits{M.}},
\oauthor{\bsnm{Kharat}, \binits{N.}},
\oauthor{\bsnm{Skau}, \binits{E.}},
\oauthor{\bsnm{Truong}, \binits{D.}},
\oauthor{\bsnm{Eren}, \binits{M.}},
\oauthor{\bsnm{Rajopadhye}, \binits{S.}},
\oauthor{\bsnm{Djidjev}, \binits{H.}},
\oauthor{\bsnm{Alexandrov}, \binits{B.}}:
pyDRESCALk: Python Distributed Non Negative RESCAL Decomposition with
  Determination of Latent Features.
\doiurl{10.5281/zenodo.5758446}.
\url{https://doi.org/10.5281/zenodo.5758446}
\end{botherref}
\endbibitem

\bibitem{eren2022general}
\begin{botherref}
\oauthor{\bsnm{Eren}, \binits{M.E.}},
\oauthor{\bsnm{Moore}, \binits{J.S.}},
\oauthor{\bsnm{Skau}, \binits{E.}},
\oauthor{\bsnm{Moore}, \binits{E.}},
\oauthor{\bsnm{Bhattarai}, \binits{M.}},
\oauthor{\bsnm{Chennupati}, \binits{G.}},
\oauthor{\bsnm{Alexandrov}, \binits{B.S.}}:
General-purpose unsupervised cyber anomaly detection via non-negative tensor
  factorization.
Digital Threats: Research and Practice
(2022)
\end{botherref}
\endbibitem

\bibitem{eren2022fedsplit}
\begin{botherref}
\oauthor{\bsnm{Eren}, \binits{M.E.}},
\oauthor{\bsnm{Richards}, \binits{L.E.}},
\oauthor{\bsnm{Bhattarai}, \binits{M.}},
\oauthor{\bsnm{Yus}, \binits{R.}},
\oauthor{\bsnm{Nicholas}, \binits{C.}},
\oauthor{\bsnm{Alexandrov}, \binits{B.S.}}:
Fedsplit: One-shot federated recommendation system based on non-negative joint
  matrix factorization and knowledge distillation.
arXiv preprint arXiv:2205.02359
(2022)
\end{botherref}
\endbibitem

\bibitem{eren2022senmfk}
\begin{botherref}
\oauthor{\bsnm{Eren}, \binits{M.E.}},
\oauthor{\bsnm{Solovyev}, \binits{N.}},
\oauthor{\bsnm{Bhattarai}, \binits{M.}},
\oauthor{\bsnm{Rasmussen}, \binits{K.}},
\oauthor{\bsnm{Nicholas}, \binits{C.}},
\oauthor{\bsnm{Alexandrov}, \binits{B.S.}}:
Senmfk-split: Large corpora topic modeling by semantic non-negative matrix
  factorization with automatic model selection.
arXiv preprint arXiv:2208.09942
(2022)
\end{botherref}
\endbibitem

\bibitem{fevotte2009nonnegative}
\begin{bchapter}
\bauthor{\bsnm{F{\'e}votte}, \binits{C.}},
\bauthor{\bsnm{Cemgil}, \binits{A.T.}}:
\bctitle{{Nonnegative matrix factorizations as probabilistic inference in
  composite models}}.
In: \bbtitle{2009 17th European Signal Processing Conference},
pp. \bfpage{1913}--\blpage{1917}
(\byear{2009}).
\bcomment{IEEE}
\end{bchapter}
\endbibitem

\bibitem{phan2008multi}
\begin{bchapter}
\bauthor{\bsnm{Phan}, \binits{A.H.}},
\bauthor{\bsnm{Cichocki}, \binits{A.}}:
\bctitle{{Multi-way nonnegative tensor factorization using fast hierarchical
  alternating least squares algorithm (HALS)}}.
In: \bbtitle{Proc. of The 2008 International Symposium on Nonlinear Theory and
  Its Applications}
(\byear{2008})
\end{bchapter}
\endbibitem

\bibitem{kim2012fast}
\begin{botherref}
\oauthor{\bsnm{Kim}, \binits{J.}},
\oauthor{\bsnm{Park}, \binits{H.}}:
{Fast nonnegative tensor factorization with an active-set-like method},
311--326
(2012)
\end{botherref}
\endbibitem

\bibitem{kim2014algorithms}
\begin{barticle}
\bauthor{\bsnm{Kim}, \binits{J.}},
\bauthor{\bsnm{He}, \binits{Y.}},
\bauthor{\bsnm{Park}, \binits{H.}}:
\batitle{{Algorithms for nonnegative matrix and tensor factorizations: A
  unified view based on block coordinate descent framework}}.
\bjtitle{Journal of Global Optimization}
\bvolume{58}(\bissue{2}),
\bfpage{285}--\blpage{319}
(\byear{2014})
\end{barticle}
\endbibitem

\bibitem{battenberg2009accelerating}
\begin{bchapter}
\bauthor{\bsnm{Battenberg}, \binits{E.}},
\bauthor{\bsnm{Wessel}, \binits{D.}}:
\bctitle{{Accelerating Non-Negative Matrix Factorization for Audio Source
  Separation on Multi-Core and Many-Core Architectures.}}
In: \bbtitle{ISMIR},
pp. \bfpage{501}--\blpage{506}
(\byear{2009})
\end{bchapter}
\endbibitem

\bibitem{fairbanks2015behavioral}
\begin{barticle}
\bauthor{\bsnm{Fairbanks}, \binits{J.P.}},
\bauthor{\bsnm{Kannan}, \binits{R.}},
\bauthor{\bsnm{Park}, \binits{H.}},
\bauthor{\bsnm{Bader}, \binits{D.A.}}:
\batitle{{Behavioral clusters in dynamic graphs}}.
\bjtitle{Parallel Computing}
\bvolume{47},
\bfpage{38}--\blpage{50}
(\byear{2015})
\end{barticle}
\endbibitem

\bibitem{moon2020alo}
\begin{bchapter}
\bauthor{\bsnm{Moon}, \binits{G.E.}},
\bauthor{\bsnm{Ellis}, \binits{J.A.}},
\bauthor{\bsnm{Sukumaran-Rajam}, \binits{A.}},
\bauthor{\bsnm{Parthasarathy}, \binits{S.}},
\bauthor{\bsnm{Sadayappan}, \binits{P.}}:
\bctitle{{ALO-NMF: Accelerated locality-optimized non-negative matrix
  factorization}}.
In: \bbtitle{Proceedings of the 26th ACM SIGKDD International Conference on
  Knowledge Discovery \& Data Mining},
pp. \bfpage{1758}--\blpage{1767}
(\byear{2020})
\end{bchapter}
\endbibitem

\bibitem{phipps2019software}
\begin{barticle}
\bauthor{\bsnm{Phipps}, \binits{E.T.}},
\bauthor{\bsnm{Kolda}, \binits{T.G.}}:
\batitle{{Software for sparse tensor decomposition on emerging computing
  architectures}}.
\bjtitle{SIAM Journal on Scientific Computing}
\bvolume{41}(\bissue{3}),
\bfpage{269}--\blpage{290}
(\byear{2019})
\end{barticle}
\endbibitem

\bibitem{mejia2015nmf}
\begin{barticle}
\bauthor{\bsnm{Mej{\'\i}a-Roa}, \binits{E.}},
\bauthor{\bsnm{Tabas-Madrid}, \binits{D.}},
\bauthor{\bsnm{Setoain}, \binits{J.}},
\bauthor{\bsnm{Garc{\'\i}a}, \binits{C.}},
\bauthor{\bsnm{Tirado}, \binits{F.}},
\bauthor{\bsnm{Pascual-Montano}, \binits{A.}}:
\batitle{{NMF-mGPU: non-negative matrix factorization on multi-GPU systems}}.
\bjtitle{BMC bioinformatics}
\bvolume{16}(\bissue{1}),
\bfpage{1}--\blpage{12}
(\byear{2015})
\end{barticle}
\endbibitem

\bibitem{lopes2010non}
\begin{bchapter}
\bauthor{\bsnm{Lopes}, \binits{N.}},
\bauthor{\bsnm{Ribeiro}, \binits{B.}}:
\bctitle{{Non-negative matrix factorization implementation using graphic
  processing units}}.
In: \bbtitle{International Conference on Intelligent Data Engineering and
  Automated Learning},
pp. \bfpage{275}--\blpage{283}
(\byear{2010}).
\bcomment{Springer}
\end{bchapter}
\endbibitem

\bibitem{kannan2016high}
\begin{barticle}
\bauthor{\bsnm{Kannan}, \binits{R.}},
\bauthor{\bsnm{Ballard}, \binits{G.}},
\bauthor{\bsnm{Park}, \binits{H.}}:
\batitle{{A high-performance parallel algorithm for nonnegative matrix
  factorization}}.
\bjtitle{ACM SIGPLAN Notices}
\bvolume{51}(\bissue{8}),
\bfpage{1}--\blpage{11}
(\byear{2016})
\end{barticle}
\endbibitem

\bibitem{koitka2016nmfgpu4r}
\begin{barticle}
\bauthor{\bsnm{Koitka}, \binits{S.}},
\bauthor{\bsnm{Friedrich}, \binits{C.M.}}:
\batitle{{nmfgpu4R: GPU-Accelerated Computation of the Non-Negative Matrix
  Factorization (NMF) Using CUDA Capable Hardware.}}
\bjtitle{R J.}
\bvolume{8}(\bissue{2}),
\bfpage{382}
(\byear{2016})
\end{barticle}
\endbibitem

\bibitem{tang2021collaborative}
\begin{botherref}
\oauthor{\bsnm{Tang}, \binits{B.}},
\oauthor{\bsnm{Kang}, \binits{L.}},
\oauthor{\bsnm{Zhang}, \binits{L.}},
\oauthor{\bsnm{Guo}, \binits{F.}},
\oauthor{\bsnm{He}, \binits{H.}}:
{Collaborative Filtering Recommendation Using Nonnegative Matrix Factorization
  in GPU-Accelerated Spark Platform}.
Scientific Programming
\textbf{2021}
(2021)
\end{botherref}
\endbibitem

\bibitem{eswar2021planc}
\begin{barticle}
\bauthor{\bsnm{Eswar}, \binits{S.}},
\bauthor{\bsnm{Hayashi}, \binits{K.}},
\bauthor{\bsnm{Ballard}, \binits{G.}},
\bauthor{\bsnm{Kannan}, \binits{R.}},
\bauthor{\bsnm{Matheson}, \binits{M.A.}},
\bauthor{\bsnm{Park}, \binits{H.}}:
\batitle{{PLANC: Parallel Low-rank Approximation with Nonnegativity
  Constraints}}.
\bjtitle{ACM Transactions on Mathematical Software (TOMS)}
\bvolume{47}(\bissue{3}),
\bfpage{1}--\blpage{37}
(\byear{2021})
\end{barticle}
\endbibitem

\bibitem{boureima2022distributed}
\begin{botherref}
\oauthor{\bsnm{Boureima}, \binits{I.}},
\oauthor{\bsnm{Bhattarai}, \binits{M.}},
\oauthor{\bsnm{Eren}, \binits{M.E.}},
\oauthor{\bsnm{Solovyev}, \binits{N.}},
\oauthor{\bsnm{Djidjev}, \binits{H.}},
\oauthor{\bsnm{Alexandrov}, \binits{B.S.}}:
Distributed out-of-memory svd on cpu/gpu architectures.
arXiv preprint arXiv:2208.08410
(2022)
\end{botherref}
\endbibitem

\bibitem{cupy_learningsys2017}
\begin{bchapter}
\bauthor{\bsnm{Okuta}, \binits{R.}},
\bauthor{\bsnm{Unno}, \binits{Y.}},
\bauthor{\bsnm{Nishino}, \binits{D.}},
\bauthor{\bsnm{Hido}, \binits{S.}},
\bauthor{\bsnm{Loomis}, \binits{C.}}:
\bctitle{Cupy: A numpy-compatible library for nvidia gpu calculations}.
In: \bbtitle{Proceedings of Workshop on Machine Learning Systems (LearningSys)
  in The Thirty-first Annual Conference on Neural Information Processing
  Systems (NIPS)}
(\byear{2017}).
\burl{http://learningsys.org/nips17/assets/papers/paper_16.pdf}
\end{bchapter}
\endbibitem

\bibitem{harris2020array}
\begin{barticle}
\bauthor{\bsnm{Harris}, \binits{C.R.}},
\bauthor{\bsnm{Millman}, \binits{K.J.}},
\bauthor{\bparticle{van~der} \bsnm{Walt}, \binits{S.J.}},
\bauthor{\bsnm{Gommers}, \binits{R.}},
\bauthor{\bsnm{Virtanen}, \binits{P.}},
\bauthor{\bsnm{Cournapeau}, \binits{D.}},
\bauthor{\bsnm{Wieser}, \binits{E.}},
\bauthor{\bsnm{Taylor}, \binits{J.}},
\bauthor{\bsnm{Berg}, \binits{S.}},
\bauthor{\bsnm{Smith}, \binits{N.J.}},
\bauthor{\bsnm{Kern}, \binits{R.}},
\bauthor{\bsnm{Picus}, \binits{M.}},
\bauthor{\bsnm{Hoyer}, \binits{S.}},
\bauthor{\bparticle{van} \bsnm{Kerkwijk}, \binits{M.H.}},
\bauthor{\bsnm{Brett}, \binits{M.}},
\bauthor{\bsnm{Haldane}, \binits{A.}},
\bauthor{\bparticle{del} \bsnm{R{\'{i}}o}, \binits{J.F.}},
\bauthor{\bsnm{Wiebe}, \binits{M.}},
\bauthor{\bsnm{Peterson}, \binits{P.}},
\bauthor{\bsnm{G{\'{e}}rard-Marchant}, \binits{P.}},
\bauthor{\bsnm{Sheppard}, \binits{K.}},
\bauthor{\bsnm{Reddy}, \binits{T.}},
\bauthor{\bsnm{Weckesser}, \binits{W.}},
\bauthor{\bsnm{Abbasi}, \binits{H.}},
\bauthor{\bsnm{Gohlke}, \binits{C.}},
\bauthor{\bsnm{Oliphant}, \binits{T.E.}}:
\batitle{Array programming with {NumPy}}.
\bjtitle{Nature}
\bvolume{585}(\bissue{7825}),
\bfpage{357}--\blpage{362}
(\byear{2020}).
\doiurl{10.1038/s41586-020-2649-2}
\end{barticle}
\endbibitem

\bibitem{MPI4PY}
\begin{barticle}
\bauthor{\bsnm{Dalcin}, \binits{L.}},
\bauthor{\bsnm{Fang}, \binits{Y.-L.L.}}:
\batitle{mpi4py: Status update after 12 years of development}.
\bjtitle{Computing in Science \& Engineering}
\bvolume{23}(\bissue{4}),
\bfpage{47}--\blpage{54}
(\byear{2021})
\end{barticle}
\endbibitem

\bibitem{2020SciPy-NMeth}
\begin{barticle}
\bauthor{\bsnm{Virtanen}, \binits{P.}},
\bauthor{\bsnm{Gommers}, \binits{R.}},
\bauthor{\bsnm{Oliphant}, \binits{T.E.}},
\bauthor{\bsnm{Haberland}, \binits{M.}},
\bauthor{\bsnm{Reddy}, \binits{T.}},
\bauthor{\bsnm{Cournapeau}, \binits{D.}},
\bauthor{\bsnm{Burovski}, \binits{E.}},
\bauthor{\bsnm{Peterson}, \binits{P.}},
\bauthor{\bsnm{Weckesser}, \binits{W.}},
\bauthor{\bsnm{Bright}, \binits{J.}},
\bauthor{\bsnm{{van der Walt}}, \binits{S.J.}},
\bauthor{\bsnm{Brett}, \binits{M.}},
\bauthor{\bsnm{Wilson}, \binits{J.}},
\bauthor{\bsnm{Millman}, \binits{K.J.}},
\bauthor{\bsnm{Mayorov}, \binits{N.}},
\bauthor{\bsnm{Nelson}, \binits{A.R.J.}},
\bauthor{\bsnm{Jones}, \binits{E.}},
\bauthor{\bsnm{Kern}, \binits{R.}},
\bauthor{\bsnm{Larson}, \binits{E.}},
\bauthor{\bsnm{Carey}, \binits{C.J.}},
\bauthor{\bsnm{Polat}, \binits{{\. I}.}},
\bauthor{\bsnm{Feng}, \binits{Y.}},
\bauthor{\bsnm{Moore}, \binits{E.W.}},
\bauthor{\bsnm{{VanderPlas}}, \binits{J.}},
\bauthor{\bsnm{Laxalde}, \binits{D.}},
\bauthor{\bsnm{Perktold}, \binits{J.}},
\bauthor{\bsnm{Cimrman}, \binits{R.}},
\bauthor{\bsnm{Henriksen}, \binits{I.}},
\bauthor{\bsnm{Quintero}, \binits{E.A.}},
\bauthor{\bsnm{Harris}, \binits{C.R.}},
\bauthor{\bsnm{Archibald}, \binits{A.M.}},
\bauthor{\bsnm{Ribeiro}, \binits{A.H.}},
\bauthor{\bsnm{Pedregosa}, \binits{F.}},
\bauthor{\bsnm{{van Mulbregt}}, \binits{P.}},
\bauthor{\bsnm{{SciPy 1.0 Contributors}}}:
\batitle{{{SciPy} 1.0: Fundamental Algorithms for Scientific Computing in
  Python}}.
\bjtitle{Nature Methods}
\bvolume{17},
\bfpage{261}--\blpage{272}
(\byear{2020}).
\doiurl{10.1038/s41592-019-0686-2}
\end{barticle}
\endbibitem

\bibitem{awan2016efficient}
\begin{bchapter}
\bauthor{\bsnm{Awan}, \binits{A.A.}},
\bauthor{\bsnm{Hamidouche}, \binits{K.}},
\bauthor{\bsnm{Venkatesh}, \binits{A.}},
\bauthor{\bsnm{Panda}, \binits{D.K.}}:
\bctitle{{Efficient large message broadcast using NCCL and CUDA-aware MPI for
  deep learning}}.
In: \bbtitle{Proceedings of the 23rd European MPI Users' Group Meeting},
pp. \bfpage{15}--\blpage{22}
(\byear{2016})
\end{bchapter}
\endbibitem

\bibitem{add1}
\begin{barticle}
\bauthor{\bsnm{Quigley}, \binits{E.}},
\bauthor{\bsnm{Holme}, \binits{I.}},
\bauthor{\bsnm{Doyle}, \binits{D.M.}},
\bauthor{\bsnm{Ho}, \binits{A.K.}},
\bauthor{\bsnm{Ambrose}, \binits{E.}},
\bauthor{\bsnm{Kirkwood}, \binits{K.}},
\bauthor{\bsnm{Doyle}, \binits{G.}}:
\batitle{“data is the new oil”: citizen science and informed consent in an
  era of researchers handling of an economically valuable resource}.
\bjtitle{Life Sciences, Society and Policy}
\bvolume{17}(\bissue{1}),
\bfpage{1}--\blpage{13}
(\byear{2021})
\end{barticle}
\endbibitem

\bibitem{add2}
\begin{botherref}
\oauthor{\bsnm{Hickey}, \binits{A.}}:
Zettabytes of data hog up space and resources
(2019)
\end{botherref}
\endbibitem

\bibitem{add3}
\begin{bbook}
\bauthor{\bsnm{Akhgar}, \binits{B.}},
\bauthor{\bsnm{Saathoff}, \binits{G.B.}},
\bauthor{\bsnm{Arabnia}, \binits{H.R.}},
\bauthor{\bsnm{Hill}, \binits{R.}},
\bauthor{\bsnm{Staniforth}, \binits{A.}},
\bauthor{\bsnm{Bayerl}, \binits{P.S.}}:
\bbtitle{Application of Big Data for National Security: a Practitioner’s
  Guide to Emerging Technologies}.
\bpublisher{Butterworth-Heinemann}, \blocation{???}
(\byear{2015})
\end{bbook}
\endbibitem

\bibitem{add5}
\begin{barticle}
\bauthor{\bsnm{Sierra}, \binits{R.G.}},
\bauthor{\bsnm{Laksmono}, \binits{H.}},
\bauthor{\bsnm{Kern}, \binits{J.}},
\bauthor{\bsnm{Tran}, \binits{R.}},
\bauthor{\bsnm{Hattne}, \binits{J.}},
\bauthor{\bsnm{Alonso-Mori}, \binits{R.}},
\bauthor{\bsnm{Lassalle-Kaiser}, \binits{B.}},
\bauthor{\bsnm{Gl{\"o}ckner}, \binits{C.}},
\bauthor{\bsnm{Hellmich}, \binits{J.}},
\bauthor{\bsnm{Schafer}, \binits{D.W.}}, \betal:
\batitle{Nanoflow electrospinning serial femtosecond crystallography}.
\bjtitle{Acta Crystallographica Section D: Biological Crystallography}
\bvolume{68}(\bissue{11}),
\bfpage{1584}--\blpage{1587}
(\byear{2012})
\end{barticle}
\endbibitem

\bibitem{add6}
\begin{barticle}
\bauthor{\bsnm{Sandberg}, \binits{R.L.}},
\bauthor{\bsnm{Huang}, \binits{Z.}},
\bauthor{\bsnm{Xu}, \binits{R.}},
\bauthor{\bsnm{Rodriguez}, \binits{J.A.}},
\bauthor{\bsnm{Miao}, \binits{J.}}:
\batitle{Studies of materials at the nanometer scale using coherent x-ray
  diffraction imaging}.
\bjtitle{JOM}
\bvolume{65},
\bfpage{1208}--\blpage{1220}
(\byear{2013})
\end{barticle}
\endbibitem

\bibitem{add7}
\begin{barticle}
\bauthor{\bsnm{Butter}, \binits{A.}},
\bauthor{\bsnm{Plehn}, \binits{T.}},
\bauthor{\bsnm{Schumann}, \binits{S.}},
\bauthor{\bsnm{Badger}, \binits{S.}},
\bauthor{\bsnm{Caron}, \binits{S.}},
\bauthor{\bsnm{Cranmer}, \binits{K.}},
\bauthor{\bsnm{Di~Bello}, \binits{F.A.}},
\bauthor{\bsnm{Dreyer}, \binits{E.}},
\bauthor{\bsnm{Forte}, \binits{S.}},
\bauthor{\bsnm{Ganguly}, \binits{S.}}, \betal:
\batitle{Machine learning and lhc event generation}.
\bjtitle{SciPost Physics}
\bvolume{14}(\bissue{4}),
\bfpage{079}
(\byear{2023})
\end{barticle}
\endbibitem

\bibitem{add8}
\begin{botherref}
\oauthor{\bsnm{Gubaev}, \binits{K.}},
\oauthor{\bsnm{Podryabinkin}, \binits{E.V.}},
\oauthor{\bsnm{Shapeev}, \binits{A.V.}}:
Machine learning of molecular properties: Locality and active learning.
The Journal of chemical physics
\textbf{148}(24)
(2018)
\end{botherref}
\endbibitem

\bibitem{add9}
\begin{barticle}
\bauthor{\bsnm{Kruglov}, \binits{I.}},
\bauthor{\bsnm{Sergeev}, \binits{O.}},
\bauthor{\bsnm{Yanilkin}, \binits{A.}},
\bauthor{\bsnm{Oganov}, \binits{A.R.}}:
\batitle{Energy-free machine learning force field for aluminum}.
\bjtitle{Scientific reports}
\bvolume{7}(\bissue{1}),
\bfpage{8512}
(\byear{2017})
\end{barticle}
\endbibitem

\bibitem{add10}
\begin{botherref}
\oauthor{\bsnm{Haghighatlari}, \binits{M.}},
\oauthor{\bsnm{Heidar-Zadeh}, \binits{F.}},
\oauthor{\bsnm{Hirn}, \binits{M.}},
\oauthor{\bsnm{Hoja}, \binits{J.}},
\oauthor{\bsnm{Isayev}, \binits{O.}},
\oauthor{\bsnm{Kondor}, \binits{R.}},
\oauthor{\bsnm{Li}, \binits{L.}},
\oauthor{\bsnm{Li}, \binits{Y.}},
\oauthor{\bsnm{Martyna}, \binits{G.}},
\oauthor{\bsnm{Meila}, \binits{M.}}, et al.:
Ipam program on machine learning \& many-particle systems-recent progress and
  open problems
(2017)
\end{botherref}
\endbibitem

\bibitem{add11}
\begin{bchapter}
\bauthor{\bsnm{Messina}, \binits{P.}},
\bauthor{\bsnm{Lee}, \binits{S.}}:
\bctitle{The us exascale computing project}.
In: \bbtitle{Proc. ACM/IEEE Conf. Supercomputing (Birds a Feather)}
(\byear{2016})
\end{bchapter}
\endbibitem

\bibitem{add12}
\begin{barticle}
\bauthor{\bsnm{Zhang}, \binits{J.}},
\bauthor{\bsnm{Xiao}, \binits{M.}},
\bauthor{\bsnm{Gao}, \binits{L.}}:
\batitle{An active learning reliability method combining kriging constructed
  with exploration and exploitation of failure region and subset simulation}.
\bjtitle{Reliability Engineering \& System Safety}
\bvolume{188},
\bfpage{90}--\blpage{102}
(\byear{2019})
\end{barticle}
\endbibitem

\bibitem{add13}
\begin{barticle}
\bauthor{\bsnm{Franke}, \binits{B.}},
\bauthor{\bsnm{Plante}, \binits{J.-F.}},
\bauthor{\bsnm{Roscher}, \binits{R.}},
\bauthor{\bsnm{Lee}, \binits{E.-s.A.}},
\bauthor{\bsnm{Smyth}, \binits{C.}},
\bauthor{\bsnm{Hatefi}, \binits{A.}},
\bauthor{\bsnm{Chen}, \binits{F.}},
\bauthor{\bsnm{Gil}, \binits{E.}},
\bauthor{\bsnm{Schwing}, \binits{A.}},
\bauthor{\bsnm{Selvitella}, \binits{A.}}, \betal:
\batitle{Statistical inference, learning and models in big data}.
\bjtitle{International Statistical Review}
\bvolume{84}(\bissue{3}),
\bfpage{371}--\blpage{389}
(\byear{2016})
\end{barticle}
\endbibitem

\end{thebibliography}
\end{document}